\documentclass[sigconf,authorversion]{acmart}

\usepackage{xcolor}
\usepackage{graphicx}
\usepackage{dirtytalk}
\usepackage{multirow}

\usepackage{CJKutf8}

\usepackage{makecell}
\usepackage{verbatim}

\AtBeginDocument{%
  \providecommand\BibTeX{{%
    \normalfont B\kern-.5em{\scshape i\kern-.25em b}\kern-.8em\TeX}}}

\acmConference[CHI '25]{CHI Conference on Human Factors in Computing Systems}{April 26-May 1, 2025}{Yokohama, Japan}

\begin{document}

\begin{CJK}{UTF8}{min}

\newcommand{\PreserveBackslash}[1]{\let\temp=\\#1\let\\=\temp}
\newcolumntype{C}[1]{>{\PreserveBackslash\centering}p{#1}}
\newcolumntype{R}[1]{>{\PreserveBackslash\raggedleft}p{#1}}
\newcolumntype{L}[1]{>{\PreserveBackslash\raggedright}p{#1}}

\title[]{``Can't believe I'm crying over an anime girl'': Public Parasocial Grieving and Coping Towards VTuber Graduation and Termination}

\author{Ken Jen Lee}
\email{kenjen.lee@uwaterloo.ca}
\affiliation{%
  \institution{University of Waterloo}
  \streetaddress{200 University Ave W}
  \city{Waterloo}
  \state{Ontario}
  \country{Canada}
  \postcode{N2L 3G1}
}

\author{PiaoHong Wang}
\email{piaohongwang@gmail.com}
\affiliation{%
  \institution{City University of Hong Kong}
  \state{Hong Kong}
  \country{China}
}

\author{Zhicong Lu}
\email{zlu6@gmu.edu}
\affiliation{%
  \institution{George Mason University}
  \city{Fairfax}
  \state{Virginia}
  \country{United States}
}

\renewcommand{\shortauthors}{Lee, Wang, and Lu}
\renewcommand{\shorttitle}{Can't believe I'm crying over an anime girl}

\begin{abstract}
{Despite the significant increase in popularity of {Virtual YouTubers} (VTubers), research on the unique dynamics of viewer-VTuber parasocial relationships is  nascent.}
This work investigates how {English-speaking} viewers grieved VTubers whose identities are no longer used, an interesting context as the \textit{nakanohito} (i.e., the person behind the VTuber identity) is usually alive post-retirement and might ``reincarnate'' as another VTuber.
We propose a typology for VTuber retirements and analyzed 13,655 Reddit posts and comments spanning nearly three years using mixed-methods.
{Findings include how} viewers coped using methods similar to when losing loved ones, alongside novel coping methods reflecting different attachment styles.
Although emotions like sadness, shock, concern, disapproval, confusion, and love decreased with time, regret and loyalty showed opposite trends.
Furthermore, viewers' reactions situated a VTuber identity within a community of content creators and viewers.
{We also discuss design implications alongside implications on the VTuber ecosystem and future research directions.}
\end{abstract}



\maketitle

\section{Introduction}

On June 8th, 2021, Kiryu Coco's audience received bad news: an announcement was made regarding Coco's graduation, which was happening in less than a month \cite{cocoGradAnnouncement}. On the day of graduation, a graduation stream was held on Coco's YouTube channel, attracting a peak concurrent viewer count of more than 500 thousand viewers \cite{dexertoCoco}. During this period, fans also gathered in various online spaces to grieve the end of a well-loved VTuber. For some fans, this might be their very first time experiencing the loss of a \textit{parasocial relationship} (PSR). For more experienced VTuber fans, this could just be another solemn reminder that no VTuber lasts forever. Indeed, at the time of writing, there have been more than 600 instances of known and recorded VTuber retirements \cite{fandomRetired}; Coco's was not the first, and definitely not the last.

VTubers are content creators using avatars to create typical online content (usually via live streams) with an identity that is an interesting mix of authenticity and performance, co-constructed by the streamer and viewers \cite{wan2024investigating}.
Prior works have explored VTubers' reasons for becoming a VTuber and their identity formation \cite{liudmila2020designing, Bredikhina2022,bredikhina2022babiniku,Chen2024}, how viewers perceive VTubers \cite{Stein2022,Lu2021}, and why fans watch VTubers \cite{Lu2021}. 
Interestingly, viewers form PSRs with the VTubers they watch and support, {even though VTubers {typically} do not disclose their real-life appearance or identity \cite{Lu2021}.}

Parasocial relationships (PSRs) are not new, and many prior works have explored PSRs and how people react to the inevitable dissolution of PSRs. 
A primary finding is how these one-sided relationships can be as significant as their two-sided counterparts; they could be just as intimate and affect people's behaviours \cite{Kowert2021em, Schramm2008}.
The widespread use of social media also resulted in the use of online spaces for parasocial grieving. 
For instance, Mou et al. \cite{Mou2023} employed a primarily data-mining-based method to compare online grieving on Bilibili between the indefinite hiatus of VTuber Kizuna AI, the death of two non-VTuber Bilibili content creators, and the death of two celebrities.
They found that unique sets of keywords were used in each context: with words like ``come back'' and ``good night'' in Kizuna AI's case, signalling hope for a potential ``rebirth'', vs. ``death'' and ``forever'' for the human celebrities.
The use of a concert as a send-off for Kizuna AI also led to the dual use of the comments section for remarks of farewell and entertainment.

This work aims to further explore the intersection of the previously discussed phenomena:  online parasocial grieving of announced VTuber retirements over multiple years. 
It differs from most parasocial grief research focusing on PSRs with celebrities or TV show characters by investigating differences in the grieving process within the VTuber context.
It also differs from prior VTuber research by focusing on the {English-speaking} audience instead of the {Chinese-speaking} audience (e.g., \cite{Tan2023, Mou2023, Lu2021, regis2023vtubers,Kim2021,Bredikhina2022,bredikhina2022babiniku}), and employing a mixed-method process to generate nuanced insights into multiple types of VTuber retirement over a larger temporal range.
{Specifically, this work investigates viewers' emotional reactions and coping methods when VTubers retire, other factors affecting viewers' reactions, and whether there is grief policing within the community.}
To do this, we analyzed 1293 posts and 12,362 comments {on Reddit related to the most frequently discussed VTuber retirements:} the announced graduation of Kiryu Coco and the announced termination of Uruha Rushia across four subreddits over almost three years using both deductive and inductive coding, followed by a statistical analysis of the coding outcomes. Quantitative analysis results were further complemented by nuanced qualitative insights.


Our empirical findings regarding viewers' emotional reactions when parasocially grieving retired VTubers show that while negative emotions like sadness, shock, concern, and disapproval dissipated over time, emotions like loyalty increased. 
We also demonstrate and discuss why expressions of love decreased in the long term despite short-term increases that replicate prior research \cite{Bingaman2022xv} and its methodological implications on qualitative parasocial grief research.
Qualitative evidence was also found for not just existing coping methods in parasocial grief \cite{DeGroot2015, Bingaman2022xv, Akhther2023xe}, but novel methods that reflect social grieving processes and different attachment styles {in the VTuber context}.
Beyond that, our findings confirm hypotheses by existing grief policing research on policing behaviours across different online spaces \cite{Gach2017} and contribute insights into cross-platform policing.
Furthermore, we contribute to research on avatar embodiment and VTuber identity co-construction \cite{wan2024investigating} through new insights on the role of collaborative content creation and viewer perceptions of VTuber reincarnations.
Last but not least, {we further discuss} the role of social media design and algorithms, followed by design implications for online spaces to better support the parasocial grieving process.


\section{Background Information}

Given that research on VTubers and online grieving within the VTuber context has just begun to develop, we provide a description of the history of VTubers and the important nuances of VTuber retirements to introduce uninformed readers to the background. 

\subsection{A Brief History of VTubers}

{The eventual emergence of VTubers can arguably be traced back to the creation of virtual idols \cite{conti2022virtual}. Almost three decades ago, in 1996, one of Japan's largest talent agencies, HoriPro, commissioned a Japanese computer graphics company, Visual Science Laboratory, to create the world's first virtual idol---Kyoko Date---a computer graphics human model voiced by multiple human voice actresses \cite{Black2008,kyokodateWiki,kyokonewspaper}. 
Kyoko participated in activities that a real-person idol would do, like releasing a music CD and participating in a radio talk show \cite{Black2008}.
Instead of investing in ``the moulding and packaging of a biological body in order to create a product which can only ever approximate an ideal,'' virtual idols were perceived as an alternative that can be designed to fit consumer tastes perfectly, without worries of potential scandals, or losses in appeal due to aging \cite{Black2008}.
Following Kyoto's footsteps, other virtual idols were created thereafter, e.g., Yuki Terai \cite{miyake2023virtual, Black2008}. 
However, none were met with the level of international success that Hatsune Miku has \cite{their2016hatsune}.
Hatsune Miku was initially created in 2007 as a visual representation of a voice synthesizer that was part of a series of software called Vocaloid, produced by Crypton Future Media \cite{leavitt2016producing}.
The widespread sharing of music created using the Vocaloid software, followed by fan-made Miku dance videos (i.e., using the MikuMikuDance software), online and offline gatherings of fans, as well as international concerts, popularized Miku globally \cite{leavitt2016producing}. These phenomena, alongside their overlaps with the manga and anime cultures, would later inspire the advent of VTubers.}

In 2016, the widely regarded pioneer of VTubers, Kizuna AI, started creating content across multiple social media platforms, including YouTube, Niconico and Twitter \cite{suan2021performing}. Her content, including videos of gameplays, skits, original music videos, and collaborations with other VTubers \cite{suan2021performing}, gained worldwide attention, garnering her more than three million subscribers on YouTube alone as of October 2023 \cite{kizunaaiyoutube}.
Major VTuber agencies like COVER Corp. (which owns the hololive production VTuber group), ANYCOLOR Inc. (which owns the NIJISANJI VTuber group) and VShojo \cite{vtubersin2023} have played a significant role ever since then; a ranking of the top 10 most popular (by watch hours) female streamers on YouTube and Twitch in 2023 include four hololive VTubers, with Usada Pekora ranking first with 29.7 million hours watched \cite{popular2023femalestreamers}.
Aside from the biggest agencies, many other companies have also launched their own VTuber(s) \cite{brandvtubers}, for example, Crunchyroll-Hime by anime streaming platform Crunchyroll \cite{crunchyrollhimenews} and GX Aura by Opera, the company behind the Opera GX browser \cite{operanews}.
Alongside the continual involvement of corporations is a flourishing indie VTuber scene, which has grown tremendously in size thanks to the increasing availability of technology and resources required to create content as a VTuber.
Statistics based on YouTube and Twitch showed that independent VTubers make up 73.9\% of all VTuber channels and 29.4\% of the 1.1 billion total hours watched of VTuber content \cite{vtubersin2023}.
{Unsurprisingly, the often anime/manga-inspired aesthetics of VTuber designs and interactions usually attract viewers already immersed in the anime/manga communities, even though recent efforts have gradually pushed VTubers into the mainstream (e.g., \cite{holododgers}).}

\subsection{VTuber Retirements}\label{sec:vtuberretirements}

\begin{figure}[ht]
  \centering
  \includegraphics[width=\linewidth,trim={4cm 4.5cm 4cm 4.5cm},clip]{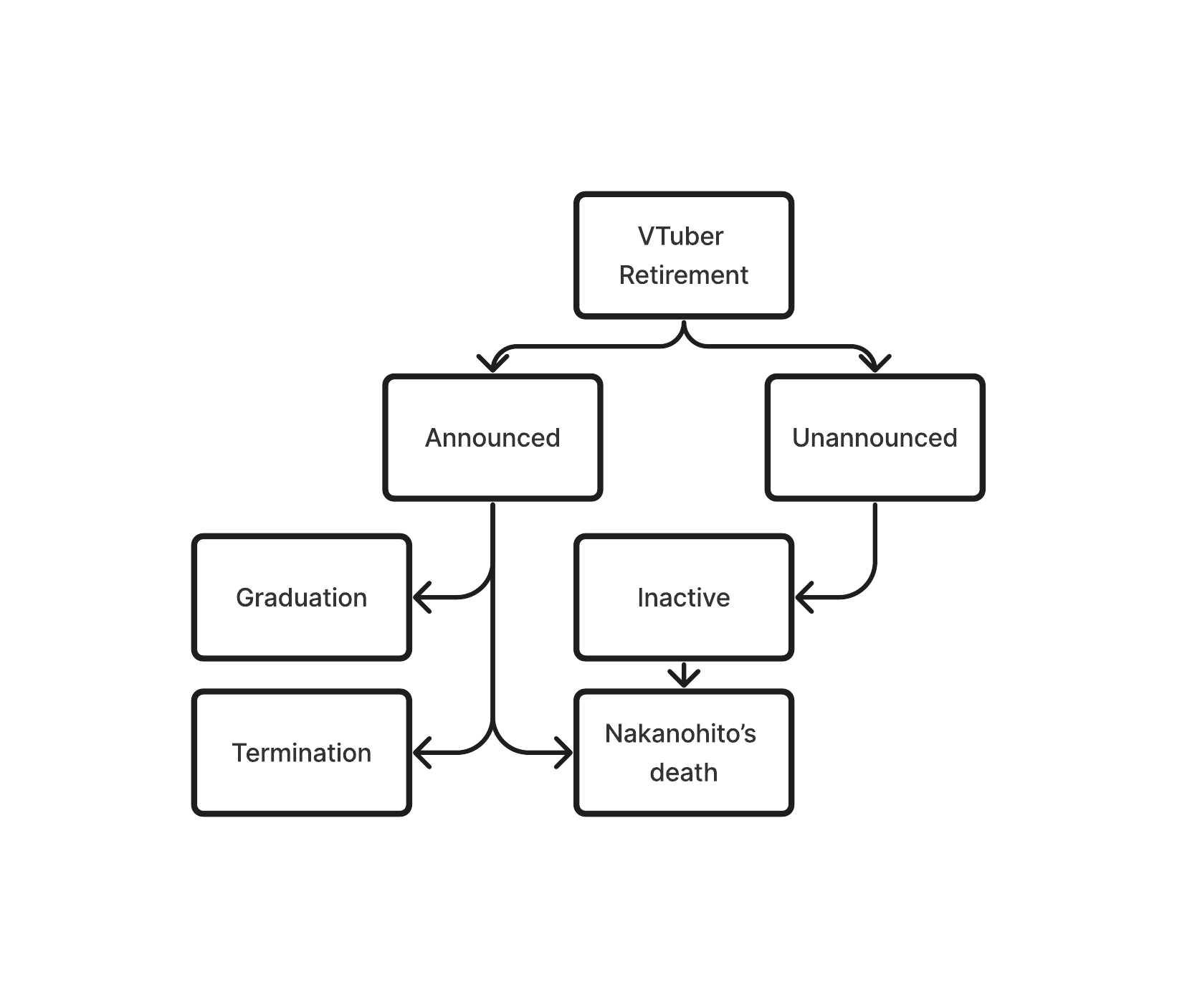}
  \caption{An overview of the types of VTuber retirements generated from the exploratory phase (Section \ref{sec:exploratory}).}
  \label{fig:retirementTypes}
\end{figure}

We propose a typology for the different types of VTuber \textit{retirement} (i.e., when a particular VTuber identity is no longer used to create content), borrowing from the Virtual YouTuber fandom wiki \cite{fandomRetired} in the way we name and categorize these types, as shown in Figure \ref{fig:retirementTypes}.
We argue that as a crowdsourced wiki, it reflects a consensus by the VTuber community and serves as a good start in understanding the complexities of VTuber retirements beyond what was described in existing research (e.g., \cite{Mou2023}).
We used observations from the exploratory phase (Section \ref{sec:exploratory}) to further inform our categorization of the main retirement types.

VTuber retirements can either be announced or unannounced. Announced retirements occur when a VTuber or another representative entity (e.g., a VTuber's agency or friend) makes an announcement regarding a VTuber's retirement. 
Unannounced retirements, in contrast, are when no announcements are made.

Announced retirements could take different forms. The first is a graduation. 
Given the overlaps between the VTuber and Japanese anime and idol \textit{otaku} cultures, specific terminologies are sometimes imported into the VTuber community. 
An example is graduation, or 
卒業
(\textit{sotsugyō}) in Japanese. 
According to Lu \cite{lu2019idolized}, the term graduation was first used in the idol industry when idols Nakajima Miharu and Kawai Sonoko left Onyanko Club, their idol group in 1986. Since then, this term has been used euphemistically to refer to members leaving their idol groups \cite{lu2019idolized}.
Within the VTuber community, graduations refer to cases of announced retirements when an announcement is made in advance that a VTuber will be ending their activities.
For agency-affiliated VTubers, this entails an amicable end to the nakanohitos' employment under their agency as a VTuber. 
Normally, graduating VTubers will hold a final farewell live stream that might include a virtual concert.
In contrast, terminations hold a more negative connotation. They are usually only applicable to agency-related VTubers and refer to cases when an agency ends its contract with a VTuber's nakanohito with immediate effect. 
Terminated VTubers usually do not get a chance to bid farewell to fans. 
Beyond that, there have also been tragic cases when a nakanohito's family or friends announced on their behalf that they have passed, hence ending their VTuber activities.

The most typical type of unannounced retirement is a VTuber becoming inactive, i.e., no longer producing any content, over long periods of time for unknown reasons. 
It is possible for a VTuber to become inactive due to their nakanohito's death without viewers ever knowing about it.

Lastly, we introduce a few terminologies when discussing VTuber retirement.
A retired VTuber's nakanohito might use another VTuber identity in the future, referred to as \textit{reincarnations}. 
Furthermore, a nakanohito's previous VTuber or non-VTuber identity is referred to as their \textit{previous life}, abbreviated as PL. 

\section{Related Work}

We present relevant existing research on multiple aspects of PSRs, including reactions to PSR dissolution, PSRs in the live-streaming context, VTubers, and online grief policing and norms.

\subsection{Research on VTubers}
Although research on VTubers is still relatively nascent, it has been receiving more attention in recent years. Bredikhina investigated VTuber identity formation by surveying 95 VTubers and found that VTubers are distinct from avatars since VTubers are directly associated with their roles as entertainers and content creators {\cite{liudmila2020designing}}. Moreover, Bredikhina observed Japanese VTubers to be more interested in pursuing idol-related activities and instances where men nakanohitos chose to embody female VTuber identities \cite{liudmila2020designing}.
In subsequent works, Bredikhina and Giard further explored the latter phenomenon and uncovered how this form of digital cross-dressing and gaining recognition as being \textit{kawaii} enable nakanohitos to indulge in their fantasies, ``act and live outside of societal pressures'' \cite{Bredikhina2022}, and live as their ``ideal self'' \cite{bredikhina2022babiniku}.
Beyond that, Chen and Hu focused on the construction of female VTuber identities and found that nakanohitos include human traits and feminine tropes idealized in the otaku culture into their identities. Moreover, they termed ``virtual breaking'' to describe the entire spectrum of actions where nakanohitos `` unveil their authentic identities that are conventionally hidden from the audience during their performances'' as VTubers \cite{Chen2024}.
Moreover, Stein et al. found that both VTuber and human live-streamers elicited similar levels of parasocial interactions, even though VTubers could be perceived as having lower human likeness \cite{Stein2022}. 

Other scholars focused on particular geographical contexts when researching VTubers.
For instance, Lu et al. \cite{Lu2021} investigated the engagement dynamics between Chinese audiences and VTubers. They found that viewers perceive VTubers to be distinct from their nakanohitos, are more tolerant of their behaviours (compared to non-VTubers), and care more about the actual content instead of the nakanohito's traits or appearance.
On the other hand, Tan examined PSRs between viewers and VTubers in China during the COVID-19 pandemic and found that certain demographics reported deeper parasocial attachments, e.g., younger viewers and students. Furthermore, stronger attachments led to better stress relief for viewers \cite{Tan2023}. 
Complementing this, Regis et al. further explored the VTuber industry in China and provided case studies on the two largest VTuber agencies in China, namely A-SOUL and VirtuaReal \cite{regis2023vtubers}. 
Examining the context of South Korean viewers, Kim and Yoo found that viewers without experience watching VTubers perceived VTubers as more professional and aesthetically appealing than non-VTubers, even though non-VTuber content was rated higher on all four subscales of the AttrakDiff questionnaire \cite{Kim2021}. 

Beyond that, some research examined the technical aspects of VTubers, be it increasing computational efficiency for live streaming as a VTuber on mobile devices \cite{Zhu2023}, building highly interactive VTubers like CodeMiko \cite{Kang2021}, or presenting an example broadcasting pipeline \cite{Shirai2019}.
More recently, Li et al. \cite{Li2023} explored the AI-VTuber phenomenon (i.e., VTubers controlled not by nakanohitos but using AI models). 
Besides VTubers, there has also been research on adjacent phenomena. 
Notably, somewhat similar to VTubers are \textit{virtual influencers} (VIs), i.e., ``computer-generated characters or avatars designed and maintained by experts and digital agencies that help brands appeal to and reach desirable target groups effectively through their digital personalities'' \cite{audrezet2023virtual}. Usually, research on virtual influencers has a larger focus on marketing and brand engagement (e.g., \cite{audrezet2023virtual,9138861,guthrie2020virtual,Zhou2023}). We further contribute to this line of research by investigating the unique phenomenon of VTuber retirement, which has been under-explored but could potentially have a profound impact on viewers.

\subsection{Parasocial Relationships}\label{sec:relwork:para}

Traditionally, PSRs are one-sided ``imaginary relationships with media figures or characters'' \cite{Hu2021}. These relationships are built off of repeated one-sided encounters by viewers with figures or characters who do not reciprocate, which is different from usual two-sided social relationships. Moreover, traditional PSRs often have broad reach (e.g., a TV movie reaching a large audience) but much more restricted access (e.g., rare chances to participate in meet-and-greets) \cite{Kowert2021em}. 
Researchers today also recognize that PSRs could be formed with other types of celebrities who are not traditional media figures or characters like athletes \cite{Bingaman2022xv}, scientists \cite{Akhther2023xe}, or live-streamers \cite{Kowert2021em}. 

\subsubsection{PSRs with Live-Streamers.}
PSRs are not quite the same in the age of more active performer-audience interactions, e.g., through live-streaming.
The availability of more channels for reciprocal communication, be it synchronous (e.g., during a live stream) or asynchronous (e.g., YouTube comments), resulted in what Kowert and Daniel Jr. \cite{Kowert2021em} termed as a ``one-and-a-half'' sided PSR.
This term reflects the potential for reciprocal interactions between streamers and their audiences, a strong sense of community affiliation intensified through co-experiences, fandom cultures, wishful identification of live-streamers as aspirational figures, high emotional engagement, and increased presence due to the relatively higher availability of the streamer and their content \cite{Kowert2021em}.
Using a mediation analysis in the context of live-streamers focused on streaming game-playing content, Lim et al. \cite{Lim2020gx} found that viewers are more likely to keep watching their favourite streamer if they have stronger PSR, and viewers' PSR strength is influenced by the amount of wishful identification and emotional engagement with the streamer.
Chen \cite{Chen2021te} used the word ``personal brand communities'' (PBCs) to describe streamers branding themselves as digital celebrities and managing relationships with viewers within such communities.
An investigation of viewers' consumer behaviour in the form of digital gifting to foster viewer-streamer-viewer PSRs in live-streaming PBCs yielded three themes: i) needs consciousness, ii) rituals and traditions, and iii) the obligations of commitment and the sharing of community cultural values \cite{Chen2021te}.
The unique dynamics in VTuber-viewer interactions and the viewers' PSRs with VTubers further motivate VTuber retirements as an interesting case to investigate PSRs in the live-streaming context.

\subsubsection{Reactions to Involuntary PSR Dissolution.}
Announced VTuber retirements are a unique type of involuntary parasocial relationship dissolution.
Researchers have investigated audiences' reactions to various types of involuntary PSR dissolutions, including imagining if one's favourite TV characters were taken off-air \cite{Cohen2003kw, Cohen2004jm}, fan reactions to the finale of popular TV shows \cite{Eyal2006ij}, TV program disruptions \cite{Lather2011fy}, and celebrity death \cite{Cohen2016vr}.
They found that just like interpersonal relationship dissolution, PSR dissolution also leads to negative reactions like sadness and anger among audiences, especially among those with stronger PSRs \cite{Hu2021}.
Similarly, Baker and Cohen \cite{Baker2023} found evidence that parasocial grieving is comparable to grieving for death in social relationships: perceived closeness, not parasociality (whether the relationship was one- or two-sided), led to differences in imagined grief responses.
Jache{\'c} \cite{Jachec2021ae} researched reactions to the tragic death of Kobe Bryant, an {NBA} basketball star, in 2020. Relevant to this work is the finding that fans honoured him in the form of eulogies, physical memorabilia left in front of the Staples Center, various works of art, and internet memes.
A content analysis of relevant Reddit posts by Bingaman \cite{Bingaman2022xv} found that sadness and shock were the most common grief-related responses to Kobe Bryant's death, and reminiscence and memorialization are common. Moreover, over time, emotional responses like sadness dissipated while other emotional responses like love increased. These temporal changes highlight the importance of studying PSR dissolution longitudinally, which we motivate further in this work.

Radford and Bloch \cite{Radford2012vn} investigated responses to the death of Dale Earnhardt Sr., a race car driver, and found that fans dealt with it through two processes: introjection and incorporation. The former allows the reliving and reinterpreting of past interactions with the deceased, in addition to the reinforcement of relevant memories, in an attempt to hold on to a relationship.
The latter, on the other hand, is when ``objects representative of the deceased are used as a means of keeping some part of that person alive'' \cite{Radford2012vn}.
Moreover, they observed all five stages of loss \cite{kubler2014grief} (denial, anger, bargaining, depression, and acceptance) in online consumer responses to the death.
A similar finding emerged from an analysis of online responses to the death of Micheal Jackson \cite{Sanderson2010ku}.
A study of online reactions on X to the death of Stephen Hawking found that emotional responses included sadness, shock, confusion, love, and longing \cite{Akhther2023xe}. Besides previously mentioned coping mechanisms like individualized tributes, reminiscence, and memorialization, Akhther and Tetteh also observed advocacy \cite{Akhther2023xe}.
Unfortunately, reactions to celebrity deaths can also be much, much darker, including depression that could even lead to suicide \cite{meyrowitz1994life}.
We argue that VTuber retirements provide a unique case to further investigate how fans react to involuntary PSR dissolution, which is the focus of the current study.

\subsection{Grief Policing and Norms}
``Grief policing'' refers to ``norm enforcement practices around grief'' \cite{Gach2017}. Analyzing relevant Facebook news articles and responses related to the death of a few celebrities, Gach et al. \cite{Gach2017} found grief policing to be ubiquitous due to commenters importing conflicting norms from various contexts, be it online (e.g., other websites/social media platforms) or offline (e.g., a funeral home).
Moreover, shared norms were ineffectively formed and enforced within Facebook posts because commenters were primarily interacting transiently with strangers in a public online space. As such, the researchers hypothesized that online spaces that are more intentional, less public, and less transient (e.g., Facebook fan pages dedicated to deceased celebrities) might lead to more effective formation and enforced norms.
Other than that, researchers also found that the algorithmic ranking of content on social media platforms played an important role in online grieving. Specifically, Krutrök \cite{ErikssonKrutrk2021} observed the formation and encouragement of generally unconventional grieving norms thanks to the use of algorithms that recommend other similar grieving content (leading to ``algorithmic closeness''), leading to perceived safe online spaces.
Unlike public Facebook or Tiktok posts \cite{Gach2017,ErikssonKrutrk2021}, we analyze VTuber-related subreddits, which could be categorized as possessing a more coherent and stronger shared identity, alongside more established norms. 

\subsection{Digital Fandom \& Identities}
Research on VTuber viewers could be seen as a type of digital fandom research. 
Research on fandoms started more than 30 years ago with the recognition that such research could generate valuable insights into identity and communications \cite{bennett2014tracing}.
The increasingly widespread availability and use of technology have been ``empowering and disempowering, blurring the lines between producers and consumers, creating symbiotic relationships between powerful corporations and individual fans, and giving rise to new forms of cultural production'' \cite{pearson2010fandom}.
Specifically, Bennett argued that technological advancements affected four aspects of fandom: communication, creativity, knowledge, and organizational and civic power \cite{bennett2014tracing}.
Specifically, social media allowed for fandom presence across multiple media platforms, alongside more direct and frequent communication between fans and their objects of fandom.
Moreover, the Internet has facilitated fans' creativity through its ability to allow fan-made creations to be shared more widely, collaborations between fans and even their targets of affection.
Such collaborative efforts also led to the creation of knowledge centers that could take the form of archive websites or fan wikis \cite{bennett2014tracing}.

Fandoms could also be crucial to the self-identity of fans \cite{williams2015post}, which could be seen as the continuous integration of external events into an `ongoing ``story'' about the self' \cite{giddens1992modernity}.
Through this lens, a PSR resolution is perceived as the fans being ``cast in the role of the rejected party,'' abandoned by the fan object. A direct result of this is the need to grief (i.e., coming to terms emotionally and cognitively) before a shift in self-identity \cite{williams2015post}.






\section{Research Questions \& Hypotheses}

Existing research mostly investigated viewers' reactions to the involuntary dissolution of PSRs with real-life celebrities or fictional characters. Some of these works focused on analyzing parasocial grieving behaviours in online spaces. We build off of previous research by investigating online grieving in a novel context: announced VTuber retirements.
In addition, methodologically, this work employs a longitudinal analysis of almost three years of data, which, though called for in prior research on online and parasocial grieving (e.g., \cite{Moyer2018}), is still nascent in existing works (e.g., analyzing only ten days \cite{daniel2017valar}, three weeks \cite{Bingaman2022xv, Akhther2023xe} or four weeks \cite{Sanderson2010ku} of online PSR grieving).

Beyond the superficial concept of content creators using avatars, the VTuber context presents various interesting nuances relevant to more general investigations of an identity's degree of fictionality and previous/afterlives, as well as the role of online spaces in parasocial grieving. Thus, we ask the following research questions:

\quad\textbullet\quad\textbf{RQ1:} What are viewers' {long-term} emotional reactions to announced VTuber retirements {within the most relevant subreddits}?

Prior works have not only identified a range of emotions (e.g., love, sadness, shock) but also how they changed over a shorter time period \cite{DeGroot2015, Bingaman2022xv, Akhther2023xe,Sanderson2010ku}. We expect similar trends in how emotions change over time and hypothesize that: 

\quad\textbullet\quad\textbf{H1:} Over time, viewers react with proportionally less negative emotions.

Moreover, based on common perceptions assigned to the different types of announced retirements within the community (Section \ref{sec:vtuberretirements}), we also hypothesize that: 

\quad\textbullet\quad\textbf{H2:} A termination elicits more confused and negative reactions than a graduation.

Beyond viewers' emotional reactions, we also investigate how viewers cope with their parasocial grief and expect to observe similar coping methods discovered by prior parasocial grief research \cite{DeGroot2015, Bingaman2022xv, Akhther2023xe}:

\quad\textbullet\quad\textbf{RQ2:} What coping methods do viewers use for their parasocial grief {and why are these methods chosen}?

\quad\textbullet\quad\textbf{H3:} Viewers of VTubers cope through various methods, including tributes, reminiscence and memorialization.

Furthermore, based on observations of ubiquitous \textit{grief policing} in public online spaces used for parasocial grieving \cite{Gach2017}, we ask the following question while paying special attention to the characteristics of the online spaces analyzed in this work:

\quad\textbullet\quad\textbf{RQ3:} How do VTuber's online fan communities enforce norms around grief on Reddit, and why?

As with any grieving process, many factors could affect viewers' behaviours while parasocially grieving and the degree of these behaviours, leading to the final RQ:

\quad\textbullet\quad\textbf{RQ4:} What contextual factors affect how viewers react to VTuber retirements?

\section{Methods} \label{sec:method}

We selected Reddit as a data source because of several of its characteristics. 
First, Reddit is a large platform used daily by more than 70 million users and contains more than 100 thousand communities, known as \textit{subreddits}, which have more than 16 billion posts and comments in total \cite{redditinc}. As of February 2024, an estimated 60.11\% of the traffic to Reddit comes from the United States, followed by 5.15\% from Canada and 4.76\% from the United Kingdom \cite{febreddittraffic}.
Second, Reddit allows users to engage with subreddits by creating posts that can contain multiple types of data (e.g.,  texts, images, videos, URLs), writing comments in posts, or reacting to a piece of content (e.g., upvoting posts or comments). 
Third, unlike other more length-limited platforms like X and Instagram, Reddit posts can have up to 40,000 characters \cite{thomas2019reddit}.
Moreover, the relative anonymity allowed while creating content on Reddit means that users can more easily share their true beliefs and emotional responses \cite{Amaya2019, thomas2019reddit}, which is crucial for our RQs.
Other than that, Reddit has not been analyzed before in existing VTuber research, which has focused more on Bilibili (e.g., \cite{Chen2024, Mou2023}), YouTube (e.g., \cite{Chen2024, Kim2021}), or Twitch (e.g., \cite{Stein2022}). As such, this work complements existing works by exploring implications due to the platform's design.

\subsection{Exploratory Phase}
\label{sec:exploratory}
While previous works on online parasocial grieving mostly selected cases based on recent events (e.g., recent celebrity deaths \cite{Gach2017, Bingaman2022xv}), we designed an exploratory phase to first build an understanding of the available Reddit data related to audience reactions to VTuber retirements. We hoped to analyze retirement(s) with the most amount of data over a longer period, instead of defaulting to recent retirements.
To achieve this, we first needed a list of retired VTubers. Fortunately, the Virtual YouTuber fandom wiki, a crowdsourced collection of wiki pages about the VTuber community, contains such a list \cite{fandomRetired}.
While there is no guarantee that this list is comprehensive, it served as a good reference regardless. 
The first author saved the list of all 594 retired VTubers as of February 2, 2024. 
Then, they went through each VTuber to build an understanding of the approximate amount of relevant Reddit discussions (i.e., discussing a VTuber's retirement) and the type of retirement.
Operationally, the former was done by searching reddit.com using the phrase ``[VTuber's name] AND (graduation OR graduate OR terminate OR termination OR retirement OR retired)'' and then examining how many posts discussing the VTuber's retirement there are in the top 30 search results and their comments, ranked based on the default ``Relevance'' setting for posts from ``All Time.''
On the other hand, the latter was done by first going through the VTuber's fandom wiki page for details like the retirement date (listed as ``Unknown'' for unannounced retirements) and details surrounding the retirement. The latter is to ascertain whether a retired VTuber experienced an announced or unannounced retirement.

Several useful insights were generated from the exploratory phase. First was the primary types of retirements, which have been presented above as important background information in Section \ref{sec:vtuberretirements}. 
Sometimes, multiple VTuber retirements were announced and discussed together, e.g., the announcement of the graduation of HOLOSTARS English's Noir Vesper and Magni Dezmond \cite{magnivesperAnnouncement}.
Furthermore, the exploratory phase provided insights into the main subreddits where VTuber-related discussions are held: r/Hololive (1.2 million members),  r/VirtualYoutubers (172 thousand members), and r/Nijisanji (95 thousand members). 
This matches the results from a search for VTuber-specific subreddits through the 5000 largest subreddits \cite{largestSubreddits}.
Outside of the 5000 largest subreddits, there also exist smaller relevant subreddits that were either dedicated to other VTuber agencies, both officially (e.g., r/PhaseConnect) and unofficially (e.g., r/hololiveEN), or specific VTubers (e.g., r/KiryuCoco).

Observations from the exploratory phase contributed to the design of the main analysis. 
{Our initial intention was to start with analyzing Reddit before including qualitative data from other platforms (e.g., Discord, YouTube). However, it was clear that the size of the final Reddit dataset (\autoref{fig:dataflow}) that the two coders needed to analyzed manually~\footnote{We made the decision to not rely on AI or other computational methods during qualitative analysis to hopefully increase our immersion in the data and hence add strength to our findings \cite{tracy2024qualitative,paulus2024minutes}.} would significantly exceed existing HCI studies using similar methods (e.g., \cite{Wu2024,Kauer2021,Kou2024,Jones2019,Gui2018}).
After taking into consideration the time needed for data analysis, the scope was limited to Reddit to build a deeper understanding of a meaningful and sizable snapshot (common within fandom research \cite{bore2013studying,bennett2014tracing,booth2013changing,williams2015post}) within the VTuber community instead of sacrificing depth for breadth.}
We also initially wanted to analyze both announced and unannounced retirements. However, a ranking of the most discussed VTuber retirements quickly revealed that announced retirements were discussed much more, making it better suited for qualitative analysis in its ability to generate deeper insights, hence this work's focus on only announced retirements. Specifically, two retired VTubers stood out in terms of the amount of relevant discussion observed from the exploratory phase (i.e., only from the top 30 search results on reddit.com): Uruha Rushia (27 relevant posts and 13130 comments) and Kiryu Coco (30 relevant posts and 10067 comments). 
Conveniently, this provided a good opportunity to compare differences between responses to a VTuber graduation (i.e., Kiryu Coco) vs. termination (i.e., Uruha Rushia), two different types of announced retirement.

\subsection{Description of Analyzed VTubers}

We provide a brief description of both Uruha Rushia and Kiryu Coco before proceeding since the data collection and analysis process are focused on them.
Both Kiryu Coco and Uruha Rushia belonged to hololive, a talent group under the VTuber management group hololive production, which is owned by COVER Corporation. It is one of the largest VTuber agencies internationally, boasting over 80 talents with over 80 million total YouTube subscribers at the time of writing \cite{hololiveabout}.
hololive consists of three groups: hololive (Japan), hololive Indonesia and hololive English. Uruha Rushia was part of hololive's third generation, while Kiryu Coco belonged to its fourth generation.
Both had large audiences exceeding a million YouTube subscribers \cite{cocoyoutubechannel,rushiayoutubechannel}.
Coco held her graduation live stream on July 1, 2021, after an announcement about three weeks prior.
She was known for her bilingual (Japanese and English) content and live stream series like \textit{AsaCoco Live News} and \textit{reddit [expletive]post review!}, which facilitated the growth of the r/Hololive subreddit. 
Rushia, on the other hand, had her termination announcement made on February 24, 2022, with immediate effect.
She spoke only Japanese and was known for her voice and having a persona that was possessive \cite{rushiajealous,rushiapossessive} and self-conscious of her avatar's characteristics \cite{rushiajokecontext}.
Coco's fans were called \textit{kiryukai} while Rushia's were called \textit{fandeads}. A more detailed description is available in Appendix \ref{app:vtuberintro}.

\subsection{Data Collection}
\label{sec:datacollection}

\begin{table*}[htbp!]
\caption{Subreddits details (as of August 2024) from which data was collected; rules that could be related to reincarnations are presented as well.}
\begingroup
\begin{center}
\label{tab:subredditsize}
\begin{tabular}{ p{.13\textwidth} p{.07\textwidth} p{.7\textwidth}}
   \multicolumn{1}{c}{Subreddit}  & \multicolumn{1}{c}{\# Members} & \multicolumn{1}{c}{Rules Related to Reincarnations}\\
   \hline
   r/Hololive & 1.2 mil. & 2. No personal, sensitive information / doxxing (This applies to talents and Redditors alike. Don't even hint at it.) \\
   & & 8. No advertising VTubers outside of hololive production (Discussion is always free though!)\\
   r/VirtualYoutubers & 187k & 7. Rules on Past and Alternate Identities\\
   & & 7a. Discussion of a VTuber's alternate or past identities must be spoiler tagged, and spoilers should be clearly labeled with what they contain. Posts must be flaired 'Alter-Ego Discussion' and may not include such info in the title. This extends to hinting about such information.\\

    & & 7b. Discussion of IRL accounts of VTubers is only allowed insofar as it's directly related to their VTuber activities.\\

    & & 7c. Doxxing is strictly prohibited.\\
   r/Rushia & 4.8k & 1. No Doxxing Rushia (This includes info about her other channels/accounts.)\\
   r/KiryuCoco & 2.7k & N/A\\
  \hline
\end{tabular}
\end{center}
\endgroup
\end{table*}

\begin{figure*}[ht]
  \centering
  \includegraphics[trim={3cm 4.5cm 3cm 4.5cm},width=\linewidth,clip]{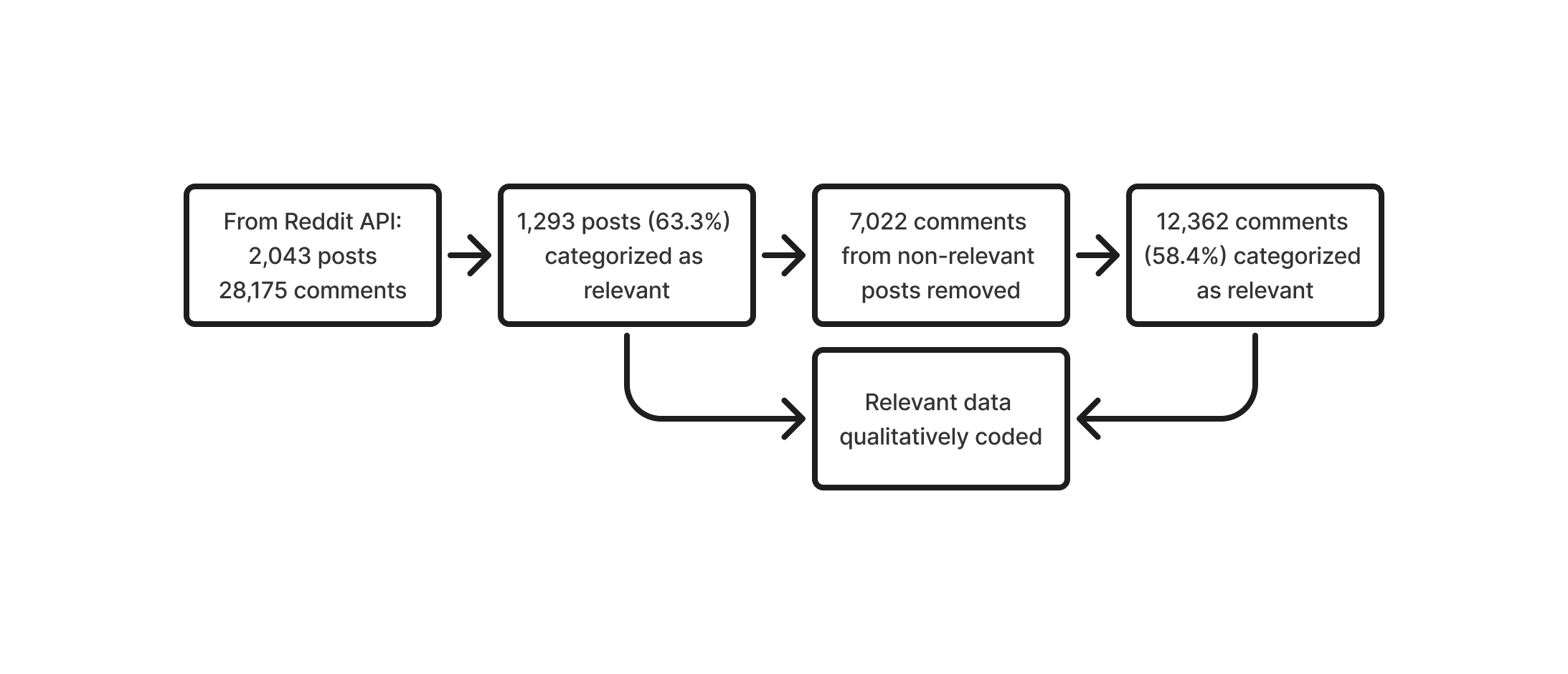}
  \caption{The data analysis process; both the categorization and coding were done manually after passing initial IRR checks.}
  \label{fig:dataflow}
\end{figure*}

Next, we performed a more comprehensive data collection phase to gather posts that are relevant to the retirements of Rushia and Coco. Informed by our observations in the exploratory phase, we chose to collect posts from r/Hololive, r/VirtualYoutubers, r/KiryuCoco, and r/Rushia (Table \ref{tab:subredditsize}).
The r/Hololive subreddit is dedicated to discussing VTubers belonging to hololive production (both Coco and Rushia were hololive VTubers), while r/VirtualYoutubers is for discussions on VTubers in general.
We also defined a post as being \textit{relevant} when:
\begin{enumerate}
    \item a post is in English; this criterion ensured our focus on the English-speaking audience
    \item a post must not have been removed or deleted
    \item a post must be primarily about the retired VTuber, judged based on a post's title and body content; i.e., a retired VTuber is not just incidentally mentioned in a post discussing other topics 
    \item a post must have been created after the VTuber's retirement was announced
    \item a post must be relevant to our research questions
\end{enumerate}

To have increased chances of searching for posts that meet the inclusion criteria, we used each VTuber's names and nicknames, which are listed on their fandom wiki entry \cite{uruharushiafandom,kiryucocofandom} as keywords to perform a search on both r/Hololive and r/VirtualYoutubers. 
This mirrors similar approaches of using relevant keywords as search terms for relevant data from social media by prior works (e.g., \cite{Akhther2023xe}).
For subreddits dedicated to the VTubers, r/KiryuCoco and r/Rushia, we queried for all the newest posts since we expect almost all posts to be relevant.
To fulfill the fourth criterion, we only collected the official retirement announcement posts \cite{cocoGradAnnouncement, rushiaRedditAnnouncement} and posts that were created after them.
These posts were collected on May 17, 2024, using RedditWarp, a Python wrapper library for Reddit's API \cite{redditwarp}. In total, 2043 posts were collected (Fig. \ref{fig:dataflow}, Table \ref{tab:datacomp}).
Following that, all top-level comments for the collected posts were retrieved as well (Fig. \ref{fig:dataflow}, Table \ref{tab:datacomp}). We only collected top-level comments since we are interested in viewers' direct reactions towards the retirements (e.g., \cite{zhang2017community,10.1145/3530190.3534848}), totalling more than 28,000 comments.

\subsection{Qualitative Analysis}
A mixed qualitative analysis process of both inductive interpretation and deductive coding using coding schemes adapted from prior works was chosen, respectively reflecting the differences and similarities between this work's investigation of viewers' reactions and parasocial grieving towards retired VTubers and prior research on reactions toward the death of celebrities and fictional characters.

\subsubsection{Deductive Codes}
Deductive coding was done using codes for two themes. The first is emotional expressions \cite{DeGroot2015, Bingaman2022xv, Akhther2023xe, Sanderson2010ku}. While some prior works used Kübler-Ross' five stages of grief (e.g., \cite{Radford2012vn, Sanderson2010ku}), we chose to use emotional expressions that were specifically observed during parasocial relationship losses instead (Table \ref{tab:deductivecodes}).
The second theme includes coping mechanisms found when parasocially grieving in public online spaces  \cite{DeGroot2015, Bingaman2022xv, Akhther2023xe} (Table \ref{tab:deductivecodes}).
We also made a few changes during the adaptation process:

\textbf{Memorialization.} As noted by DeGroot and Leith \cite{DeGroot2015}, the act of creating posts and comments about a VTuber's retirement itself is a form of memorialization. As such, we gave memorialization a more specific definition based on observations from prior works: the common focus on one's ``positive characteristics'' \cite{DeGroot2015} or ``remarkable things'' \cite{Akhther2023xe}.

\textbf{Creative Tributes.} Among the most relevant works, only Akhther and Tetteh \cite{Akhther2023xe} had a code for individualized tributes. However, anything written as a form of memorialization could technically be categorized as tribute as well, based on tribute's definition \cite{cambridgetribute}. To mitigate this overlap, we adapted this code to focus on the primary aspect highlighted in their work \cite{Akhther2023xe}: the various creative outputs made as a tribute.

\begin{table*}[htbp!]
\caption{Deductive codes used during the analysis. 
}
\begingroup
\begin{center}
\label{tab:deductivecodes}
\begin{tabular}{ p{.09\textwidth} p{.12\textwidth} p{.6\textwidth} p{.09\textwidth} }
   \multicolumn{1}{c}{Theme}  & \multicolumn{1}{c}{Code} & \multicolumn{1}{c}{Adapted Definition} & \multicolumn{1}{c}{References}\\
   \hline
   \multirow{9}{*}{\makecell{Emotional\\Expressions}} & Sadness & Explicit or implicit expressions of emotional pain towards a VTuber's retirement. & \cite{DeGroot2015, Bingaman2022xv, Akhther2023xe}\\
   & Shock & Expressions of surprise and disbelief towards a VTuber's retirement. & \cite{DeGroot2015, Bingaman2022xv, Akhther2023xe}\\
   & Longing & Expressions of missing a retired VTuber. & \cite{DeGroot2015, Bingaman2022xv, Akhther2023xe}\\
   & Love & Expressions of adoration, affection, and gratefulness towards a retired VTuber. & \cite{DeGroot2015, Bingaman2022xv, Akhther2023xe}\\
   & Confusion & Expressions of uncertainty towards the circumstances surrounding a VTuber's retirement. & \cite{DeGroot2015, Bingaman2022xv, Akhther2023xe}\\
   & Disapproval & Angry or disappointed disapproving expressions towards a retired VTuber or the circumstances surrounding a VTuber's retirement. & Code ``critical'' \cite{Sanderson2010ku}\\
   \hline
   \multirow{6}{*}{\makecell{Coping \\Mechanisms}} 
   & Creative Tributes & Creative works made as tributes to a retiring/retired VTuber. & \cite{Akhther2023xe} \\
   & Reminiscence & The sharing or consumption of a retiring/retired VTuber's content and the sharing of personal memories related to the VTuber. & \cite{DeGroot2015, Bingaman2022xv, Akhther2023xe, cambridgereminiscence}\\
   & Memorialization & To help others remember, or show that one remembers, a retiring/retired VTuber's memorable things. & \cite{DeGroot2015, Bingaman2022xv, Akhther2023xe, cambridgememorialize}\\
  \hline
\end{tabular}
\end{center}
\endgroup
\end{table*}

\subsubsection{Inductive Codes}

Besides deductive coding, inductive interpretation was also employed. 
A codebook was initially created as the coders familiarized themselves with the dataset during the data categorization and IRR process (Section \ref{sec:IRR}). Further improvements were made during the coding process.
The final inductive codebook (Appendix \ref{app:codebook}) includes 14 inductive codes, four of which are new emotion codes: respect \cite{Drummond2006}, loyalty \cite{connor2007loyalty,connor2018loyalty}, concerned \cite{borkovec1983preliminary}, and regret \cite{connolly2002regret,coricelli2007brain}. 
Constant comparisons and memoing were also performed to aid in generating insights.

\subsubsection{Inter-Rater Reliability}
\label{sec:IRR}
Existing guidelines on inter-rater reliability (IRR) \cite{McDonald2019} suggest a good fit for this work to employ IRR measurements. Specifically, in this work, the codes were intended to be used as part of the final results themselves by comparing the codes' frequencies temporally, not just during the coding process. Moreover, the data was coded by more than one coder (the first two authors), hence motivating the use of IRR statistics. 
In this work, IRR statistics were calculated in four scenarios: i) ensuring IRR when determining a post's relevance based on its title and body texts, ii) ensuring IRR when determining a comment's relevance, iii) ensuring IRR when coding data deductively using the codes in Table \ref{tab:deductivecodes}, and iv) ensuring IRR when coding data inductively (Appendix \ref{app:codebook}).
Moreover, following Hallgren's recommendations \cite{Hallgren2012}, both Siegel and Castellan’s kappa \cite{siegel1957nonparametric} and Byrt et al.’s kappa \cite{byrt1993bias} were calculated for each scenario to correct for any bias and prevalence problems respectively.

For the first scenario, 30 posts were randomly sampled and assigned as either relevant or irrelevant (based on the criteria presented above) independently by the two coders. Kappa statistics were then calculated. If unsatisfactory, disagreements were discussed and resolved, and another 30 posts were randomly sampled. This process repeated until a satisfactory IRR was obtained ($kappa > .8$ \cite{landis1977measurement,krippendorff2018content}). Thirty comments were randomly sampled per round from relevant posts in a similar process for the second scenario. Both posts and comments used identical inclusion criteria (Section \ref{sec:datacollection}).
All posts and comments used in the third and fourth scenarios were selected from a pool of posts and comments that were previously judged as relevant. This is to clearly separate between categorizing a post's or comment's relevance and how they are coded.

To achieve IRR for the deductive codes (Table \ref{tab:deductivecodes}), 10 posts and 10 top-level comments from those posts were sampled each round for coding until a satisfying IRR was achieved. 
Unlike prior works that assigned each unit of analysis (a post or comment) only a single emotion code (e.g., \cite{Sanderson2010ku}), we believe that having the flexibility to assign multiple codes (e.g., a comment expressing both confusion and shock) allows for improved analysis. 
As such, the kappa statistics were calculated for each code separately as a binary variable for each unit of analysis. 
A similar process was used for the inductive codes; additional rounds were also performed whenever changes were made to the codebook.

The final dataset consisted of 1293 posts (63.3\% relevance) and 12362 comments (58.4\% relevance), which were all coded using the deductive and inductive codes.

\subsection{Quantitative Analysis}
To understand how the viewers' emotions changed over time, a binomial regression model was built for each emotion as the outcome variable, which is a boolean value for each relevant post or comment.
Two predictors were included in the model.
The first is the amount of time between when the retirement was announced and when the post or comment was created. 
This is done by subtracting the Unix timestamp of the official post announcing a VTuber's retirement from the Unix timestamp of when the post/comment was created. 
Divisions were then made to so that the difference has a unit of days passed. 
For a fair comparison, the time range of Coco's data was limited to the time range of Rushia's data.
Before building the model, this continuous predictor was centred to allow the main effect and interaction terms to be interpreted meaningfully \cite{schielzeth2010simple}.
The second predictor is a binary variable for whether a particular post/comment was about Coco; Rushia's data was assigned \textit{false}.
The interaction term between the two predictors was only added if the sample size was sufficient. 
{The inclusion of quantitative analyses is intended as a form of methodological triangulation for increased reliability \cite{denzin2017research}.
Moreover, these models are more accurate than alternatives like rank order comparisons of qualitative codes \cite{Humble2009} to study temporal trends.}

\subsection{Statements of Positionality}
Acknowledging the positionality of the authors may be important given that this research involves a
qualitative process \cite{Fusch2018} of selecting and interpreting data. 
The first author was first exposed to VTubers through Kizuna AI's YouTube videos in 2017 and later became a more consistent viewer of VTuber-related content since mid-2020.
Generally, the author prefers to watch VTuber content by browsing Holodex.net~\footnote{Holodex is a fan-made website that indexes VTuber-related live streams, videos, and video clips: https://holodex.net/}, or those recommended by YouTube, without exclusively supporting anyone/a few particular VTuber~\footnote{This is referred to as \textit{DD}, an abbreviation of the Japanese phrase 
誰でも大好き
\cite{hologlos_dd,mdkj_dd} that means to like anyone, as opposed to having an \textit{oshi}: one or more VTuber(s) a viewer supports in particular \cite{hologlos_dd}.}
For the past four years, he has also lurked in several VTuber-related subreddits.
He has experienced the retirements of multiple VTubers that he watched casually and was inspired to pursue this research through his observations of how VTuber audiences have reacted online over the years.
In this sense, the author identifies himself as an opportunistic complete-member-researcher of the group studied who does not commit to the full range of behaviours by the studied group \cite{adler1987membership}, {also referred to as a scholar-fan within fandom research \cite{larsen2011fandom}}. 
The second author was an outside member and provided different perspectives of the VTuber community. The third author has conducted several research projects related to VTubers or VTubing. This distinction in the authors' experiences with the VTuber community was helpful for maintaining objectivity when interpreting the data in this research. 

\subsection{Ethics Considerations}

Ethical implications are especially important for research using social media data \cite{Sanderson2010ku}. 
In this work, all data analyzed were available public; no data was collected from non-public groups or forums. 
To better protect the anonymity of the Reddit users whose data was quoted for this work, we also employed a moderate level of disguise based on Bruckman's recommendations \cite{Bruckman2002}.
This meant not disclosing the author of any quote used and employing sufficient rewording to avoid being located via online searches \cite{Reagle2022}.
However, we deemed the disclosure of the subreddits analyzed as necessary to present comparisons between different online spaces.
All authors confirmed with their institutions' ethics boards that the methodology used in this work is exempted from an ethics review. 

Care was also applied regarding the alternate identities of a VTuber's nakanohito. While norms within the community regarding this vary across different online spaces (Section \ref{sec:disconlinespaces}), we adopt a more conservative approach: the existence of a nakanohito's other identities was mentioned if necessary, but their names and any associated identifying information were not disclosed.

\section{Results}

\begin{figure*}[ht]
  \centering
  \includegraphics[width=\linewidth,clip]{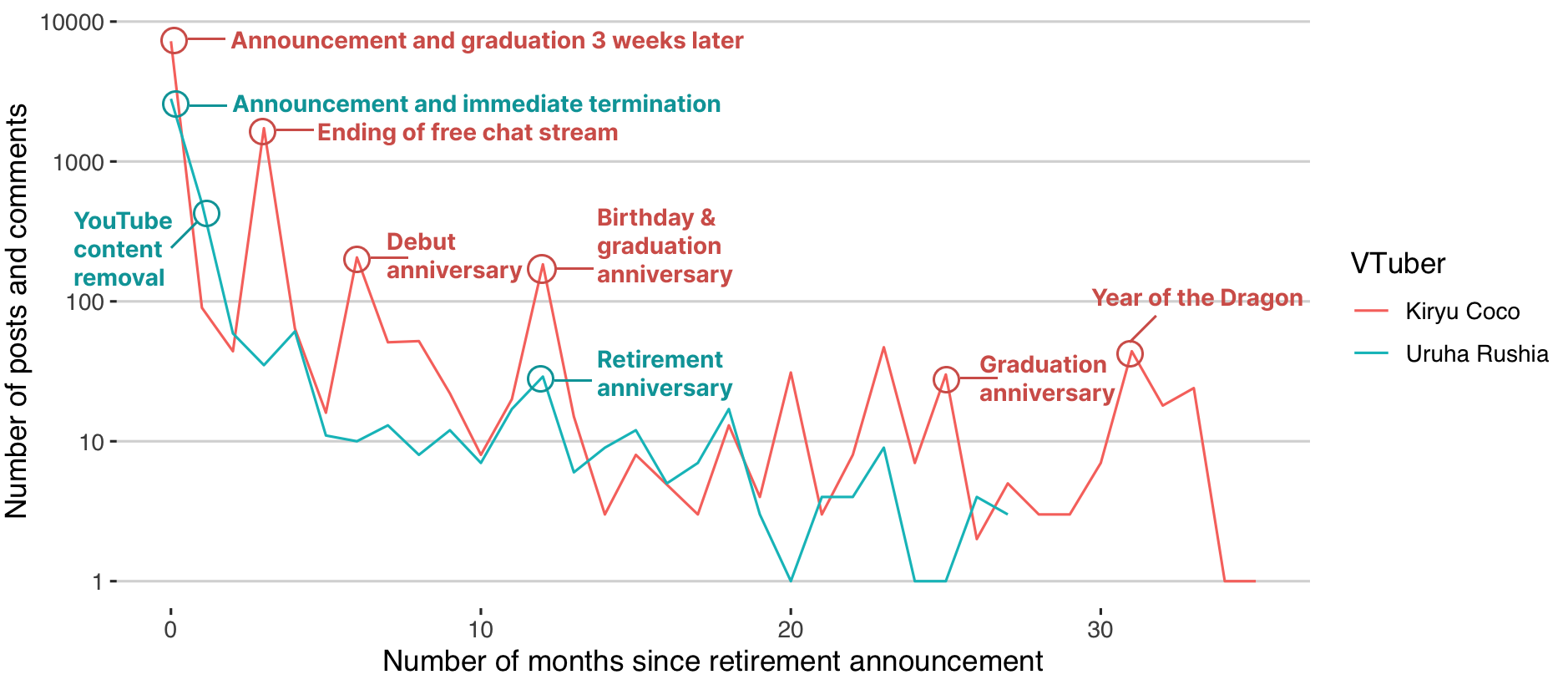}
  \caption{The number of relevant posts and comments posted over time for each analyzed VTuber.}
  \label{fig:numberdataovertime}
\end{figure*}

In general, we observed increased numbers of relevant posts and comments during notable events (Fig. \ref{fig:numberdataovertime}). Beyond the obvious (e.g., anniversaries), a huge bump was also observed at the end of Coco's free chat stream--VTubers commonly have a stream scheduled into the far future on their YouTube channel so that viewers (sometimes even the VTubers themselves) can chat in the stream's live chat. A few viewers also engaged with the subreddits when the Year of the Dragon arrived since Coco's natural form is a dragon in her lore. Other bumps were usually when a relevant post attracted numerous relevant comments, e.g. when a dedicated fan wrote a reflection about their experience with Rushia's termination.

\subsection{Viewers' Emotional Reactions (RQ1)}




\textbf{Sadness.} Sadness was expressed in many ways beyond just saying one is ``sad.'' 
As a form of emotional pain, sadness was expressed through words like being ``hurt'' or in ``pain.''
Another common expression is ``pain \textit{peko},'' referencing a verbal tic popularized by Usada Pekora, another hololive VTuber. 
Several viewers also reported physical manifestations due to their sadness, including not being able to fall asleep, feeling physical headache and chest pain, and most commonly, crying (e.g., ``Let me go cry in the shower now.''). 
To imply crying, many comments either used the metaphor of rain (``it's a terrible day to rain'', ``my roof is leaking'') or referred to the process of onion-cutting (``there are ninjas cutting onions now'').
Sometimes, this sadness can become ``unbearable'' and result in depression (e.g., ``I am so depressed now it's crazy''), leading to more extreme forms of expressions like indicating possible intentions to self-harm.
The quantitative analysis found sadness to dissipate slowly over time ($\beta = -2.87e-3, p < .001, OR = 9.97e-1 (95\% CI: 9.96e-1, 9.98e-1)$). 
Moreover, responses to Coco's retirement were less sad on average ($\beta = 0.31, p < .001, OR = 1.36 (95\% CI: 1.22, 1.52)$) and faded away faster ($\beta = -7.06e-3, p < .001, OR = 9.93e-1 (95\% CI: 9.91e-1, 9.95e-1)$).
Viewers felt sadder when Rushia was terminated for two reasons: both Rushia and the viewers ``can't even say goodbye,'' since she left without being able to host a final goodbye live stream, unlike Coco, and the removal of all content from Rushia's YouTube channel from public access: ``It’s even sadder since her channel videos will not remain.''
In contrast, Coco viewers appreciated Coco being given a graduation stream and how fans were informed a few weeks before: ``her lead up to today's graduation stream helped soothe much of the pain.''

\textbf{Shock.} It is unsurprising that many viewers felt surprised by Coco's and Rushia's retirement. After all, both VTubers had a large audience and ranked among the most superchatted YouTube channels globally when retiring.
Their retirement felt ``surreal'' as viewers were ``not ready for this.''
This inability to accept the retirement announcement as being real was expressed as either having a bad dream (``please let this nightmare end'') or perceiving the announcement as some form of joke (``please say sike'', ``must be a late April's fools right?''). 
A few viewers even reported having goosebumps, feeling nauseous, numb (``Just numb...I can't believe it occurred.''), dizzy (``I feel dizzy, did not see this coming''), and even uncontrollable trembles (``I am actually shaking now, it's hard to accept the idea that my daily life has suddenly worsened.'').
Overall, viewers found Coco's retirement to be less shocking ($\beta = -0.54, p < .001, OR = 0.58 (95\% CI: 0.52, 0.66)$). 
In fact, the suddenness of Rushia's termination resulted in desperate pleas (``nonono please not Rushia'') and many comments simply stating ``what the [expletive]'' in disbelief. 
Regardless of which VTuber, shock faded over time ($\beta = -0.01, p < .001, OR = 0.99 (95\% CI: 0.98, 0.99)$).

\textbf{Concerned.}  Viewers felt less concerned for Coco's nakanohito ($\beta = -2.74, p < .001, OR = 0.06 (95\% CI: 0.04, 0.10)$).
A VTuber's termination could imply that things did not end amicably with their agency (Section \ref{sec:vtuberretirements}). Coupled with possible signs that Rushia's nakanohito was experiencing high levels of stress prior to her termination, many viewers first thought of her mental wellness when her retirement was announced. 
E.g., ``Legit more worried about her mental health,'' ``I just want to know she's okay.''
Similar to previous emotions, feelings of concern faded with time ($\beta = -0.07, p < .001, OR = 0.93 (95\% CI: 0.90, 0.95)$), in no small part thanks to viewers knowing a nakanohito is well from their reincarnation post-retirement (``she is doing okay these days'').

\textbf{Disapproval.} Since both VTubers analyzed belonged to an agency, disapproving remarks were targeted at both the nakanohito and the agency.
For example, Rushia's termination announcement stated contract violation as a reason, leading to comments that blamed Rushia's nakanohito (``she should not have broken her contract, that's it'').
However, most disapproving remarks were targeted at how hololive handled both Coco's and Rushia's retirement, be it perceiving the word ``graduation'' as a euphemism for forceful removal (``Why can't you just say she was fired?''), or 
how Rushia was not given a final stream (``at least let her have a final stream''), among other reasons, speculative or not. 
In several instances, this led to viewers explicitly stating their intent to stop watching hololive's VTubers altogether (``I'm out, I'm sticking with [another VTuber agency]'').
Unsurprisingly, viewers had less disapproving responses to Coco's retirement ($\beta = -2.36, p < .001, OR = 0.09 (95\% CI: 0.04, 0.19)$).
Overall, disapproval decreased with time ($\beta = -0.02, p < .001, OR = 0.98 (95\% CI: 0.97, 0.99)$).

\textbf{Regret.} Viewers' sense of regret often stemmed from their inability to still interact with a VTuber post-retirement. They reported regretting not showing more support to a retired VTuber either financially through YouTube channel membership and superchats or verbally and not knowing them earlier (``I wished I learned about you even earlier'').
Quantitatively, even though regret increased gradually over time in general ($\beta = 3.56e-3, p < .01, OR = 1.00 (95\% CI: 1.00, 1.00)$), Coco's viewers had a decreasing likelihood of expressing regret with time ($\beta = 5.33e-3, p = .01, OR = 0.99 (95\% CI: 0.99, 1.00)$).

\textbf{Confusion.} A sudden retirement announcement led to not just disbelief but also confusion as well due to not knowing what led to the announcement (``So confused, what happened? I thought things were going well'').
Some viewers who were perhaps less familiar with VTubers felt confused about certain terminology (``graduation from where?'', ``is she dying?'').
Others thought Coco left the company when her graduation announcement was released and were confused when she continued making content for the final few weeks before her graduation date (``I thought Coco has left?'')
Unsurprisingly, confusion dissipates with time as well ($\beta = 1.96e-3, p = .05, OR = 9.98e-1 (95\% CI: 9.96e-1, 1.00)$). Viewers also felt less confused when Coco retired ($\beta = -1.02, p < .001, OR = 0.36 (95\% CI: 0.27, 0.48)$) as Coco herself had a few weeks before her graduation to interact with her viewers and clarify the situation. 

\textbf{Longing.} Viewers not only missed a retired VTuber (``My heart has a dragon-shaped hole'') but also their interaction with others in the agency (``I miss kanacoco moments,'' referring to interactions between Coco and fellow hololive VTuber Kanata) since a retired VTuber usually loses their ability to have collaborative live streams with other VTubers in the agency.
Besides that, there is no strong quantitative evidence that the amount of longing changed over time or that the likelihood of a longing remark being created differed between Coco and Rushia.

\textbf{Respect.} Respect was most commonly expressed by the salute emoji ``o7''. When Coco's graduation was announced, several viewers expressed their acceptance of the announcement as they ``respect Kaichou's [Coco's nickname] decision.''
A few viewers showed their respect to Coco either through memes depicting how they intended to wear formal attire to attend her graduation stream or actually doing that in real life.
Overall, respect did not change with time. However, in Coco's case, viewers were much more likely to write posts and comments showing respect on average ($\beta = 2.20, p < .001, OR = 9.01 (95\% CI: 7.53, 10.87)$). This likelihood even increased with time ($\beta = 3.69e-3, p < .001, OR = 1.00 (95\% CI: 1.00, 1.01)$).
This does not necessarily mean Rushia is perceived as deserving less respect but that viewers chose to express their positive emotions differently, especially since Coco's perceived larger direct influence on the {English-speaking} VTuber community on Reddit was discussed frequently.

\textbf{Loyalty.} A common expression of loyalty was to pronounce the immortality of a retired VTuber's fanbase, e.g., ``Kiryukai forever!'', ``Fandead forever.'' 
Before graduating, Coco mentioned the possibility of her returning after 500 years, leading to viewers counting down the days even years after her retirement (``only 498 years left!'').
Moreover, viewers were also loyal to the nakanohitos (``I will support her regardless of where she goes''), leading to questions asking about a retired VTuber's reincarnations.
Loyalty increased over time ($\beta = 1.91e-3, p < .001, OR = 1.002 (95\% CI: 1.001, 1.003)$). Viewers responded with higher loyalty to Coco on average ($\beta = 0.76, p < .001, OR = 2.13 (95\% CI: 1.78, 2.56)$), but it increased at a slower rate with time ($\beta = -1.47e-3, p = .02, OR = 9.99e-1 (95\% CI: 9.97e-1, 1.00)$).

\textbf{Love.} Other than explicit expressions of love (e.g., the heart emoji), many viewers expressed their gratefulness for having had the opportunity to watch a retired VTuber and its wonderful memories or for being ``pulled down the rabbit hole'' of VTubers and hololive by the retired VTuber.
The word ``arigathanks''---a combination of thanks and \textit{arigatou} that emerged from a trend of humorously making ridiculous multilingual combinations popularized by Coco---was used commonly when thanking Coco.
We also observed viewers expressing their sincere thanks to the VTubers for helping them live through very difficult times in their lives.
Surprisingly, love decreased slowly with time ($\beta = -2.47e-3, p = .02, OR = 9.98e-1 (95\% CI: 9.95e-1, 9.99e-1)$). While viewers were more likely to react to Coco's retirement with more love on average ($\beta = 1.15, p < .001, OR = 3.16 (95\% CI: 2.71, 3.71)$), it decreased quicker with time ($\beta = -4.44e-3, p < .001, OR = 9.96e-1 (95\% CI: 9.93e-1, 9.98e-1)$).
A separate model limited to the first three weeks of data showed love increasing over time instead ($\beta = 0.06, p < .01, OR = 1.06 (95\% CI: 1.02, 1.10)$).


\subsection{Viewers' Coping Methods (RQ2)}
Below we present observations on both coping methods that have already been studied and other novel methods as well.

\textbf{Creative tributes.} We observed a plethora of tribute types, including memes, drawings (digital and traditional), comics (often with a manga style), cosplays, animations, music (music covers, original tribute music, remixes of original songs by the retired VTuber), as well as in-game tributes (e.g., a monument in Minecraft, custom mods to bring a retired VTuber into a game).
Interestingly, memes were not always humorous and instead expressed rich emotions of wholesomeness, sadness, love, etc.
Many also created clips (either one or multiple moments from a VTuber's live streams) to honour Coco and Rushia.
These clips often displayed clear creative decisions, e.g., adding visual effects onto a live stream moment and selecting moments that best represent a VTuber's personality and identity.
There were several notable tributes that involved considerable fan collaboration, like displaying compilations of fanarts in public advertisement spaces, creating a song cover with hundreds of fans singing along, or a scrapbook with screenshots from a VTuber's live streams over the years.
Another unique example was when viewers gathered in r/Rushia to collaborate on drawing a tribute pixel art on r/place.

\textbf{Reminiscence.} Viewers reminisced their various personal memories about the retired VTubers. Many reminisced how they came across Coco's and Rushia's content, which became their gateway into VTubers (``many of us fell down this rabbit hole after watching Coco's clip in our YouTube recommendations''). Moreover, many were exposed to Coco and Rushia not through their live streams but through fan-made clips. These clips often contain English subtitles, allowing viewers to understand the VTubers' Japanese content. 
For some, watching their live streams became a daily/weekly routine (``I have always watched her morning streams before school''), even if it meant learning Japanese to watch live streams live without English translations (``even though Rushia streams around 4 AM, I woke up every day without fail to watch them ... I even started studying Japanese'').

Other than that, viewers also reminisced by either re-watching Coco's and Rushia content (``I am re-watching her debut to relive my memories''), or sharing them with others. The latter is done either through sharing a live stream's URL directly, a clip with notable moments, or with a series of live stream screenshots, paired with English subtitles if the intention is to share what the VTuber said.
That said, a few viewers reported not being able to watch a retired VTuber's content, since it makes them too emotional (``I actively avoided replaying the video to not cry''). Although surprising initially, the first author found himself also avoiding content about Minato Aqua, another hololive VTuber who retired in August 2024, for similar reasons.

\textbf{Memorialization.} Viewers memorialized various aspects of Coco and Rushia. Both were remembered for being hardworking and fantastic entertainers (``Coco was a brilliant entertainer'') who were caring towards their viewers and fellow VTubers (``her sincere kindness is rare'').
Both VTubers were also acknowledged as playing an essential role in introducing {English-speaking} viewers to hololive. Beyond that, many viewers also missed Rushia's ability to both have ``EVA-01'' screams and also an ``angelic voice'' and the resulting ``gap \textit{moe}.''~\footnote{``Gap \textit{moe}'' refers to when characters become more likeable due to contrasting characteristics \cite{shibuya2019male}.}

On the other hand, Coco was particularly acknowledged for her influence on the {English-speaking} VTuber community beyond hololive. Her bilingual content and ``multicultural background'' were particularly praised as ``a bridge between the Engish-speaking viewers and other Japanese-speaking hololive members,'' as well as having ``united the kaigai and JP-niki'' (i.e., the overseas and Japanese viewers).  Her \textit{reddit [expletive]post review!} live streams popularized the r/Hololive subreddit, since viewers were motivated to post memes for Coco to react on stream. Moreover, her success with the {English-speaking} audience was seen as a motivating factor for hololive English's formation (``Coco planted the seeds of holoEN''). Beyond hololive, Coco was also perceived as having ``brought the VTuber industry to the West,'' even inspiring some to become VTubers themselves (``I was inspired by Coco kaichou when I started VTubing''). 

Besides that, both retired VTubers were memorialized through their numerical achievements. These include view count milestones (e.g., when their original music video reaches \textit{n} million views or how Coco's graduation stream became the second most viewed VTuber live stream had over 480,000 live viewers, a record surpassed only by Minato Aqua's graduation stream in August 2024 \cite{minatoaquagrad}) and how much they received in YouTube superchats.

\textbf{Other observed coping methods.} We also observed viewers coping in other ways:
\begin{enumerate}
    \item Buying merchandise: Viewers wanted to buy merchandise to remember the VTuber.
    \item Imagining hypothetical scenarios: When Coco was retiring, many viewers hoped that she would either remain in hololive in a managerial role or join another VTuber agency.
    \item Interacting with other viewers: We observed Rushia's viewers sharing their reflections in r/Rushia and ``getting my [their] feelings to a community open to listening,'' which others found relatable. Besides that, some fans also organized an unofficial gathering celebrating Coco's Birthday after she graduated.
    \item Sharing references: Viewers would share third-party content about Coco and Rushia, e.g., tweets made by other VTubers sharing how they felt about the retirements or things that made viewers think of them, like a photo of a cloud that looked like a dragon.
    \item Refusal to leave: A few viewers refused to leave Coco's graduation stream and live chat stream even after it ended, hoping for an Easter egg. In Rushia's case, a few fans stayed on her YouTube channel page to witness her videos being removed from public access live, even counting down to the moment from 24 hours prior in r/Rushia.
    \item Finding substitutes: A few viewers wanted recommendations on other VTubers that were similar to a retired VTuber, their reincarnation being the main substitute. In Coco's case, viewers also hoped for other hololive VTubers to take over her \textit{reddit [expletive]post review!} live stream series. 
    \item Taking a break: When viewers get overwhelmed by a VTuber's retirement, they may take a break, e.g., not interact with r/Hololive or watch VTubers for some time.
\end{enumerate}

\subsection{Grief Policing and Moderation (RQ3)} 
The r/Hololive subreddit was locked down (no new posts allowed) for some time after both retirement announcements and during Coco's graduation stream. 
Viewers were largely welcoming towards these actions, seeing them as necessary to prevent spam (``it's a good move to lock the subreddit'').
That said, a few viewers felt the subreddit was locked down for too long and preventing viewers from creating posts on other hololive related events.
In Rushia's case, an official appreciation thread was created in r/Hololive where viewers could share tributes in the comments instead of through separate posts. 
While many viewers appreciated an avenue to express their thanks for Rushia, others were more cynical of the moderators' intention.
Beyond that, many comments asked others to ``be understanding'', not blame any party blindly, and not discuss the retirements in the live chat of other hololive VTubers' YouTube live streams unless the topic was brought up by the VTuber.
A few comments also hoped for those visiting from r/all to ``chill''.
Other than that, we observed several posts in r/VirtualYoutubers noting how they posted in that subreddit due to not having sufficient karma to post in r/Hololive, a behaviour that r/VirtualYoutubers moderators warned against to prevent karma begging.

\subsection{Contextual Factors Affecting Viewers' Reactions (RQ4)}
Here, we present qualitative evidence of factors that could have contributed to how viewers reacted to the retirements. 

\textbf{Prior experiences with retirements.} Several comments noted how a viewer either had no or considerable experience in dealing with VTuber retirements. 
While prior experiences might help with dealing with VTuber retirements (``if it's your first graduation stream, remember the nice memories and cherish the VTubers we have now''), that was not the case more than often: ``I cannot take this, not this again'', ``I've went through a few graduations, but none felt this heavy'', ``my 8th graduation and it feels the same.''

\textbf{Dedication.} Viewers differed in how dedicated they were as fans to a retired VTuber, but even casual fans were emotionally affected by the retirements (``even my friend who is way more of a casual fan was devastated'', ``I never watched Coco much but I still cry like a toddler...''). Moreover, they appreciated the retired VTubers all the same (``she is not my oshi but I enjoy her clips'', ``I was not a fandead but enjoyed having her in hololive'').

\textbf{Parasociality.} While all data analyzed in this study are signs of parasocial grieving, a portion of viewers reported signs of deeper parasociality. For them, experiencing a retirement felt like ``the death of a fictional character'' or the nakanohito, or losing ``a relative'', ``a family member'', ``a friend'', or ``someone close''. 
A few viewers also admitted to being \textit{gachikoi} (falling in love with the retired VTuber identity).
Often, these viewers are clearly aware of their parasociality (``I avoid getting too parasocial...'', ``don't know if that's pathetic'' or ``healthy'') and that the nakanohitos are alive when Coco and Rushia retired, surprising even themselves with their emotional reactions.



\textbf{r/all} Multiple posts related to both retirements in r/Hololive received significant engagement and appeared in r/all (described by Reddit as ``Today's top content from hundreds of thousands of Reddit communities.'' \cite{rallsubreddit}). This resulted in curiosity and confusion about what VTubers are and what a graduation/termination is (``from r/all, I have no clue what this is''). While some offered their condolences, others considered it ``creepy'' and did not understand viewers' reactions (``The streamer is not dead...why is the tone here like a funeral?'').
A few viewers, in turn, provided the necessary context (``to r/all, thank you for checking out hololive...to summarize...''). There were instances when viewers pinged a specific Reddit user who often helped with explaining what hololive is. 

\textbf{Reincarnations.} Several posts and comments hinted at both Coco's and Rushia's reincarnations in various ways while still adhering to the rules (Table \ref{tab:subredditsize}). Example phrases include being able to still ``hear her'', or describing a retired nakanohito's ``another lifetime'', ``next life'', ``alt identity'', ``new form'', ``another alias'', etc. A few viewers hinted at their reincarnations' characteristics without stating any names, while others explicitly named them in r/VirtualYouTubers, r/Rushia and r/KiryuCoco. And while many viewers ``know how to search for'' the reincarnations, there were several instances discussing how viewers gradually stopped watching them as they ``are not the same.'' Viewers pointed out differences in the type of content created and how their reincarnations no longer collaborate or interact with other hololive VTubers, even if their personalities might remain identical.

\textbf{Archival.} Being able to still access a retired VTuber's content was important for many viewers, expressed through the many questions about whether the content will remain public post-retirement. Since Rushia's content was removed from her YouTube channel, many viewers gathered (mostly in r/Rushia) to discuss where and how backups were made before the removal. Sometimes, viewers would collaborate on archiving different portions of the content on various cloud platforms, which are frequently taken down due to copyright infringement.
A few viewers also discussed the ethics of making backups of paid content (e.g., YouTube content exclusive for paying channel members) available public.





\noindent\textbf{Summary of Findings.}
The analysis revealed viewers' emotions changed over time: sadness, shock, concern, disapproval, confusion, and love decreased over time, while regret and loyalty showed the opposite; respect and longing displayed no major changes.
Viewers' reactions to Coco's retirement also exhibited several differences compared to the entire dataset.
Findings revealed coping methods that include tributes, reminiscence, memorialization and other novel methods.
Instances of grief policing were observed as well.
Furthermore, we found factors like viewers' prior experiences with VTuber retirements and dedication level to not have a major effect on their reactions. 
We also observed reactions from non-viewers visiting from r/all, various expressions of parasociality, and discussions of the retired VTubers' reincarnations and the archival status of their content.

\section{Discussion}

Here, we discuss the support for \textbf{H1-3} and how our findings compare with existing parasocial grief research. 
We further consider how a shared identity influenced grief policing behaviour, the role of interactions between content creators in the co-construction of VTuber identities, how different online spaces resulted in different viewer reactions and design implications for various aspects of helping viewers' grief and move on.

\subsection{Viewers' Emotional Reactions (RQ1)}
Viewers were less likely to respond with most negative emotions (sadness, shock, confusion, feeling concerned, disapproval) over time, hence supporting \textbf{H1}. Interestingly, while love increased in the first three weeks, replicating Bingaman's findings \cite{Bingaman2022xv}, it decreased over the longer term. 
We argue that this is not necessarily because viewers felt less love for the VTubers, but that they express it through words that represent adjacent positive emotions, potentially explaining the increase in loyalty.
These differing patterns motivate the need for a coding process with more emotion codes to better capture the various positive and negative emotions; in our work, coding only for love would not have generated these nuanced insights.
The distinct short- and long-term emotional changes also motivate the analysis of data with larger temporal ranges in future works.
Regardless, the gratefulness viewers showed towards the retired VTubers for being in their lives reflects similar sentiments when parasocially grieving a TV show character's death \cite{Gerace2024}.
Furthermore, Coco's retirement elicited responses that were less sad, shocking, concerning, and disapproving and more respectful, loyal and loving. As such, \textbf{H2} is strongly supported; viewers react more negatively to terminations than graduations.
Making the retirement announcement weeks before the retirement gave viewers time to process it while still being able to interact with Coco. The final graduation live stream also provided closure to many viewers. Moreover, the graduation live stream, which included a final chat with other hololive VTubers and a concert performance, could be seen as rituals that helped relieve the viewers' grief \cite{meyrowitz1994life, Gerace2024}.
In contrast, the suddenness of Rushia's termination that resulted in more negative reactions reflects similar findings when reacting to a sudden loss of a loved one \cite{Krychiw2018}.
Besides that, unlike findings by Mou et al. \cite{Mou2023}, viewers often used the word ``forever'' to show their intention to stay a fan of a retired VTuber forever or to forever support a retired VTuber's nakanohito.

\subsection{Viewers' Coping Methods (RQ2)}
Tributes, reminiscence and memorialization were all observed, supporting \textbf{H3} {and aligning with existing research on coping mechanisms for PSR dissolution \cite{DeGroot2015, Bingaman2022xv, Akhther2023xe}.
Moreover, our observations contribute towards fandom research that found how personal stories of people becoming fans play an important role when meeting other fans \cite{duffett2013understanding}.
Particularly, these stories also play a central role when reminiscing after a PSR dissolution.}

We also observed novel coping methods, from purchasing merchandise and sharing references to imagining hypothetical scenarios. 
The sharing of lengthy reflections in online spaces and validating responses from other viewers reflect similar practices when grieving the loss of a loved one \cite{Matthews2019,swartwood2011surviving}.
Interestingly, this behaviour might indicate a secure attachment style, while viewers who chose to take a (permanent) break from watching VTubers in fear of experiencing another hurtful retirement could indicate a fearful-avoidant attachment style \cite{levy2011attachment,mikulincer2010attachment}.
The gathering of fans on r/Rushia to stay on Rushia's YouTube channel's page while her content was being removed draws parallels to players gathering in a game's final hours \cite{consalvo2012happens}.
These similarities extend to the search for substitute VTubers and content \cite{consalvo2012happens}.

{We argue that underlying many coping methods is a desire by viewers to maintain their self-identity as it relates to the retired VTubers‘ and their communities.
For instance, many tributes involve remixing (e.g., in a compilation clip) or recreating (e.g., fanart) memorable moments of a VTuber. Creating or consuming these tributes could be a way of reliving these experiences, evidenced by comments about fond memories of viewers experiencing these moments for the first time.
The recalling of how viewers first became fans could be seen as a reflection of their self-identity and an effort to hold on to it.
Similarly, other coping methods like buying merchandise and a refusal to leave a retired VTuber's online profile page that is about to be changed point to the viewers' wish to keep their current self-identity intact and perhaps even a refusal to move on and shift their self-identity accordingly.}

\subsection{Grief Policing (RQ3)}

Policing comments were not ubiquitous since shared norms were relatively well enforced within subreddits dedicated to VTuber viewers, aligning with hypotheses by prior research \cite{Gach2017}. 
The stronger shared identity also led to policing messages often using a friendly, instead of critical, tone.
While algorithmic ranking could allow parasocial grievers to find similar grieving posts in a non-topic-specific feed and facilitate the formation of a safe grieving space \cite{Gach2017}, that was not needed in subreddits that are already dedicated to the VTuber community.
In fact, the import of conflicting norms by non-viewers visiting popular r/hololive posts from r/all due to Reddit's algorithmic ranking was a source of policing.
That said, the overwhelming number of viewers compared to non-viewers in these posts drowned out the latter's messages \cite{erzikova2020drowning}.
We also observed cross-platform grief policing: viewers asking others to not cope by questioning other hololive VTubers while they stream on YouTube.
This is not surprising since the subreddits are more centralized spaces for viewer-viewer interactions compared to YouTube comments.

\subsection{Contextual Factors Affecting Viewers' Reactions (RQ4)}
Surprisingly, while dedicated fans were obviously affected by the retirements, more casual viewers also reported being affected to a somewhat similar degree. 
This finding, however, might be rather specific to hololive VTubers. Specifically, casual viewers also felt sad because they perceived a retired VTuber as being a part of the larger hololive family.
This sense of community between hololive's VTubers and viewers, akin to a one-and-a-half-sided PSR \cite{Kowert2021em}, highlights the social nature of the hololive fandom \cite{Coppa2014}, which could have a stronger collective identity than the general VTuber community.
As such, we hypothesize that less dedicated viewers are more affected when it comes to VTubers who either have a smaller, tightly-knit audience or belong to a VTuber group perceived by viewers as being close friends.

Just as surprising is how viewers with more experience with VTuber retirements were not much less affected.
This suggests that even after experiencing their favourite VTuber graduating, viewers might find another VTuber(s) that they like and build a parasocial relationship over time, explaining why they get just as hurt even after going through multiple retirements. That said, experienced viewers also learned to balance their parasociality, embrace the impermanence of VTubers and cherish the interactions with whomever they are currently supporting.
We leave the discussion of another factor--how well-known a nakanohito's reincarnation is among viewers--to the next part.

\subsection{VTubers \& nakanohitos}
Observations that viewers are sad even though they know the reincarnation of a retired VTuber's nakanohito and can continue to watch them hints at the complex nature of VTuber identities.
Using the lens of self-presentation theory, a VTuber's collection of previous live streams on their social media profile (e.g., YouTube channel) represents a hybrid performance-exhibition online space \cite{Hogan2010} since it exhibits live stream footage in which VTubers perform through their interactions with viewers and other collaborating content creators (VTubers or not).
As such, their identity is not only co-constructed by the nakanohito and the viewers \cite{wan2024investigating} but also interactions with collaborators.
In other words, a VTuber's reincarnation will not have an identical identity if the identity co-construction process differs due to changes in the nakanohito's desired representation, the viewers, or the collaborators. 
This could explain why viewers reported perceiving reincarnations as feeling different even if the nakanohito tried to portray a similar identity.
Particularly, collaborations between their previous lives and other hololive VTubers became a crucial part of the retired VTuber's identity that could not be replicated in their reincarnation due to the lack of similar collaborations. 

Our findings also align with prior VTuber research \cite{Lu2021} showing how viewers who view VTubers as friends beyond a fictional character care more for the nakanohito than the VTuber identity's consistency, opposing the Hyperpersonal Model of computer-mediated communication \cite{walther1996computer, walther2011theories}.
This was especially salient when viewers displayed deep concern for Rushia's nakanohito when she was terminated and when viewers announced their intention to support the nakanohitos' future activities beyond their retired VTuber identity.






\subsection{Online Spaces for Fandoms}
\label{sec:disconlinespaces}
The analyzed subreddits could be categorized into three specificity levels: all VTubers (r/VirtualYoutubers), hololive VTubers (r/Hololive) and specific VTubers (r/Rushia, r/KiryuCoco). 
Compared to r/Virtualyoutubers, r/Hololive users were more dedicated to hololive VTubers specifically; they might not watch non-hololive VTubers much, just like r/Virtualyoutubers users who might not know hololive well.
As such, reactions on r/Hololive towards hololive VTubers' retirements are generally more passionate due to a stronger shared identity than in r/Virtualyoutubers.
This passion and the use of many fandom terminologies and jargon might have worsened the confusion and sometimes even criticism from non-VTuber viewers from r/all, reflecting similar challenges met by other subreddits \cite{jones2019r}.
Reddit's r/all recommendation algorithm exposed many to the VTuber community and definitely plays an essential role in introducing VTubers to new viewers but also possibly strengthening existing misconceptions about VTubers among non-viewers (via arguments with r/Hololive users within nested comments) sometimes. 
In contrast, r/Rushia and r/KiryuCoco are more private (e.g., less likely to appear in r/all) and dedicated to preserving memories of the retired VTubers where viewers can share their tributes and thoughts with others whom they perceive as being more relatable, characteristics shared by other smaller subreddits \cite{Hwang2021}.
Posts containing long reflections and questions that are extremely VTuber-specific 
were found almost exclusively in these subreddits.
While viewers were more indirect and careful when referring to the retired VTubers' reincarnations in r/Hololive, they could do so more liberally in r/Virtualyoutubers using the \textit{spoiler} tag. In contrast, the reincarnations' identities were disclosed much more freely in r/Rushia and r/KiryuCoco. 
As such, depending on the rules and perceived intimacy of an online space, viewers might interact differently, even when discussing the same topic.
Regardless, the recognition of viewer-VTuber parasocial relationships in these subreddits mitigates disenfranchised grief, which is unexpressed grief related to a relationship unrecognized by society \cite{doka1989disfranchised, Gerace2024}.

Also relevant is YouTube's recommendation algorithm. A large number of comments discussed how they started watching hololive VTubers after being recommended a clip on YouTube. Multiple hololive VTubers went viral during the pandemic period through fan-made clips and edits, introducing new viewers to the many unique VTubers within hololive. 
As such, the importance of fans, fan-created content and the online spaces that facilitate their engagement with the community cannot be overstated, mirroring fandom research highlighting technology's role in empowering fans to ``engage in the networked, participatory behaviours ... in a diverse array of fannish activities'' \cite{Coppa2014}.
{This reflects the non-existence of contradistinction between production and consumption as various online spaces facilitate the creation and sharing of fan-created content \cite{mckee2004tell}.}

{Also relevant is the spread of the community across multiple online spaces. Besides the previously discussed cross-platform grief policing, we also observed instances of Reddit users urging others to stay off X for some time post-announcement to avoid viewer reactions that might be overly emotionally charged or contain misinformation. 
It is clear that these users perceived a difference in the content and interactions within the VTuber community across different social platforms, possibly indicating a balkanization process, i.e., the separation of a community into homogeneous, polarized sub-communities \cite{Kaylor2019}.
However, we argue that these differences are instead a natural result of differences in how various platforms are designed \cite{Kwon2023}: profile-based interactions on X often resulted in content like a VTuber's upcoming stream schedule or fanarts that a VTuber likes being shared there. On the other hand, the comment nesting of Reddit allows for more diverse and in-depth discussions of various VTuber-related matters, while more personal relationships within the community are often facilitated by Discord features like two-way voice chats.
In contrast, VTuber-related videos might be shared on platforms like YouTube and Twitch, with viewer comments primarily targeted towards the content creator instead of other viewers.
As such, we observed a trend of viewers using multiple platforms to fulfill different needs while sharing platform-specific information with other platforms, e.g., cross-posting a VTuber's announcement from X or their YouTube community tab to Reddit, or welcoming Reddit users into specific Discord servers to further collaborate on fan projects. }

\subsection{Copyright: Archives \& Tributes}
{More than mere entertainment, watching a retired VTuber's content could have become an important part of a viewer's identity both within the larger VTuber community and beyond-- be it a well-established daily habit, or by embodying the values exemplified by the VTuber through their content.
As such, one common worry that viewers had when the respective retirement announcements were made concerns the availability of a VTuber's content for future consumption post-retirement. 
This is not surprising; the internet is an ``enduring ephemeral'' \cite{chun2008enduring}, YouTube might be available for the foreseeable future, but specific YouTube content, such as a retired VTuber's videos and live streams, could be removed at any time.
As observed by De Kosnik \cite{de2021rogue}, even though digital archiving tools might be largely accessible, the key lies in the collaborative efforts of ``techno-volunteers'' to sustain online archives.
The efforts by viewers in archiving a retired VTuber's content observed in this work qualifies as what De Kosnik terms ``rogue'' community archives
\cite{de2021rogue}, created and maintained by dedicated fans who want to be able to still consume a VTuber's content post-retirement, especially if they perceive future consumption to fulfill a certain purpose, be it nostalgia, escapism, or the maintenance of a PSR. 
As such, archiving could indeed be a ``powerful emotional'' experience \cite{de2021rogue}; we argue that it is the personal value assigned to a VTuber's content by a fan that determines their priority when archiving: we observed fans who only cared about archiving a certain portion of content, e.g., only song covers and originals, ASMR videos, or collaborative live streams with a specific group of VTubers.
Importantly, we observed unofficial subreddits being used as online spaces to facilitate a collaborative archival process in a never-ending tug-of-war between fans who want to preserve content that has been or might be, removed and DMCA takedown requests by VTuber agencies, mirroring similar struggles by other communities (e.g., groups trying to preserve video games that are no longer reasonably available \cite{vidgamepreserve}).
Unfortunately, we observed such takedown requests resulting in viewers feeling alienated from the VTuber agency or the larger VTuber community.

Fan-created derivative works, i.e., tributes, represent another area where fan interests might be in conflict with the interests of copyright holders \cite{de2021rogue,milne2024fanfiction}.
However, researchers have increasingly recognized the role of fan works not just as a sign of a brand's popularity \cite{brown2009harry,brown2010selling} but also as a form of marketing.
Specifically, engaging with fan works could increase consumption intent, hence strengthening a brand \cite{milne2024fanfiction}.
We argue that producing tributes also strengthens a viewer's (i.e., a fan-creator) sense of belonging and identity within the VTuber community, hence suggesting that fan works could strengthen the brand loyalty of both fan-creators and other fans engaging with these works.
One way copyright holders have tried to balance the pros and cons of fan works is to publish official guidelines, e.g., Hololive production's Derivative Works Guidelines \cite{hologuidelines}.
Such guidelines contributed to a thriving group of fan-creators who shared their emotions and experiences with others also coping with a VTuber's retirement via creative tributes.
Our observations suggest that when brands show their viewers a willingness to reach a middle ground (e.g., having official derivative work guidelines and an official and moderated space to share tributes), viewers reciprocate through increased engagement and support.}




\subsection{Comparisons with Prior Research on Chinese-speaking Viewers}

As Sapir puts it, language ``powerfully conditions all our thinking about social problems and processes'' \cite{edward1951selected}. While our methodology does not allow for a precise understanding of the cultural backgrounds of the studied population, we believe our focus on English-speaking viewers, instead of the Chinese-speaking viewers focused on by several existing studies (e.g., \cite{Tan2023, Mou2023, Lu2021,regis2023vtubers,Kim2021,Bredikhina2022,bredikhina2022babiniku})~\footnote{Although VTubers arguably originates from Japan, the authors found fewer publications written in English focused on Japanese-speaking viewers.}, brings with it certain cultural implications.
For example, compared to Lu et al.'s findings \cite{Lu2021}, our observations found viewers generally perceiving VTubers as live streamers embodying VTuber avatars instead of anime characters coming alive. 
Accordingly, while both works found a sense of company as a motivation to watch VTubers, we saw no evidence of VTuber content being perceived as ``following every episode of an anime'' \cite{Lu2021}. 
We also did not observe any mention of VTubers' ``inclinations to solicit virtual gifts from viewers,'' a dynamic that could be more prevalent in China \cite{Lu2021}. 
Moreover, the familiar manner in which many viewers reminisced the retired VTubers as close friends (albeit parasocial) seems to contrast prior findings that viewers felt more distanced to the VTubers due to their avatar and persona \cite{Lu2021}.
This closer sense of distance did not seem to take away from how much attention was paid to the actual content compared to a VTuber's characteristics \cite{Lu2021}.


\subsection{Design Implications}
We discuss several design implications pertaining to online spaces that might be used by the VTuber or other communities for parasocial grieving. 

\subsubsection{Anonymity}
{The importance of anonymity when reacting to a PSR dissolution was clearly displayed in this work. 
We observed viewers expressing their honest emotions and thoughts, often leaving themselves vulnerable.
Unfortunately, the confused and sometimes negative reactions by those visiting from r/all are a sign that the VTuber culture is not yet well understood by those unfamiliar with it or similar cultures (e.g., in the anime and manga community).
Some viewers even expressed explicitly the need to hide their sadness in real life from those around them for fear of being misunderstood or judged, {aligning with prior research on increased self-disclosure due to anonymity \cite{10.1145/2858036.2858414}.}
As such, we believe that the more public an online space is, the more important it is to provide viewers with anonymity while they grieve, {since allowing for more self-disclosure is crucial to healthy grieving (e.g., lower shame and depression levels \cite{LeviBelz2023}).}
Accordingly, viewers might be more willing to share identity-related details or unpopular but honest reactions in less public spaces like smaller subreddits (e.g., r/Rushia) and VTuber-focused Discord servers \cite{Hwang2021}. 
}

\subsubsection{Supporting Parasocial Grieving}
Ideally, an online space dedicated to parasocial grieving should be well-known enough so that viewers know of its existence but not enough for viewers to feel disenfranchised grief due to conflicting norms by non-viewers~\cite{doka1989disfranchised, Gerace2024}.
Providing moderators with post-specific options to opt-out from appearing in more public feeds (e.g., r/all on Reddit) could help.
Providing non-viewers with background information could also aid in mitigating conflicting norm policing within a community that they do not fully understand.
Setting up barriers to participating in such spaces is another possible solution. Examples include requiring a minimum amount of karma to post in a subreddit or having private channels in a Discord server.

\subsubsection{Supporting Tribute Creation}
Given the prevalence of tribute creation {both within the VTuber community (e.g., in the form of memes \cite{Lu2021}) and} as a coping method, it is crucial to support viewers' desire to do so. 
While we observed online spaces like subreddits and Discord servers facilitating viewers in collaborating when planning and executing larger tribute projects, several problems persisted.
First, inartistic viewers, often lacking the skills to create tributes alone, might want to contribute to larger collaborative efforts but do not know how or if it is possible.
{Existing research on expert-novice collaborations and crowdsourcing could be borrowed to mitigate this~\cite{Venkatagiri2019,Belghith2022}. A simple example is how a few large tributes we observed were broken down into parts that novices could contribute to, e.g., writing a thank-you message.}
Second, viewers might miss tribute-projects-related posts due to how posts are ordered in the feed, partially mitigated by new posts about a particular project being created at intervals.
Options to track available tribute projects viewers can participate in, and any skill requirements could help resolve this, e.g., a pinned post in a subreddit listing all fan projects.
{Besides that, our observations suggest that brands (e.g., VTuber corporations) should not let copyright hinder brand performance \cite{milne2024fanfiction}; having guidelines that allow viewers to engage with the brand more actively through tributes not only allows a more healthy coping process but could also increase viewers' willingness to keep supporting other VTubers in the group going forward. }

\subsubsection{VTuber Reincarnation Disclosure}
The popularity of both Coco and Rushia meant their reincarnations were relatively well-known among viewers, especially compared to other VTubers with much smaller audiences. 
For viewers who would like to watch the reincarnations but were unaware of their existence, learning about their identities and watching them might help them move on from grieving their previous lives' retirement.
In contrast, unwantedly learning about reincarnation identities could be immersion-breaking.
To balance these needs, the reincarnation disclosure norm of online spaces could be disclosed (e.g., Table \ref{tab:subredditsize}) and filtering options (e.g., spoiler tags) could be made available to help users decide whether to engage or avoid them.
Providing viewers with options to either ask about a retired VTuber's reincarnation to dedicated viewers with such information or announce their desire to learn about it so that others might tell them through more private means is also helpful.

\subsubsection{VTuber Recommendations}
{We observed viewers facing difficulties when trying to find other VTubers to watch, which could prolong the process of moving on from a VTuber's retirement. 
Often, viewers seek recommendations not just based on content type (e.g., type of games played) but more intangible characteristics like personality and dynamic with viewers, 
as well as practical considerations like language, stream schedule and platform (e.g., YouTube vs. Twitch).
Unfortunately, viewers might not be able to verbalize their preferences precisely. 
While VTuber-focused content delivery platforms like holodex.net are a big step in the right direction, more advanced systems that better elicit viewer preferences could lead to better recommendations beyond superficial characteristics (e.g., which group a VTuber belongs to, a VTuber's design and aesthetics).
A possibility would be a platform that collects crowdsourced descriptions of VTuber personalities and dynamics, with an interface that suggests best-matched VTubers based on the personalities and dynamics of the VTubers a viewer watches most frequently. 
{Many other possibilities could be borrowed from user-centric recommendation system research, e.g., eliciting viewer preferences by first recommending a wide variety of VTubers \cite{pu2008user}, using an interactive exploration and filtering process \cite{Loepp2014}.}
We argue that {effective recommendations} could help reduce the amount of required change in viewers' self-identity after the retirement of their favourite VTubers hence softening its emotional impact.
}

\subsection{Limitations}
While the popularity of hololive and particularly Coco and Rushia provided us with sufficient data to generate interesting insights into viewers' reactions over time, most VTubers have a much smaller audience size who might not even have announced retirements.
Based on the exploratory phase, these instances often leave fans wondering if they are doing okay, if and when they will be back, or if they have started activities under another VTuber identity, presenting an interesting research direction.
Future work could also investigate parasocial grieving for independent VTubers who are not agency-affiliated.
Similarly, our work does not investigate VTubers who do not use YouTube primarily (e.g., Twitch) and other viewer interactions not happening on Reddit (e.g., Discord, X).
Moreover, our categorization of VTuber retirements (Fig. \ref{fig:retirementTypes}) might be incomplete, and we encourage future research to further investigate this.

Another limitation related to Reddit as a data source is how Reddit users are ``more likely to be male and younger than the general population'' \cite{Amaya2019} {with a ``strong U.S. and Anglophone bias'' \cite{Hintz2022}}. 
Other than that, performing research using Reddit data has become more difficult due to API access changes \cite{redditapiannounce}, resulting in tools like Pushshift (a popular~\footnote{The publication about Pushshift~\cite{baumgartner2020pushshift} has more than 1,000 citations according to Google Scholar at the time of writing.} tool used to ``analyze large quantities of reddit data''~\cite{pushshiftFAQ}) no longer being accessible for researchers \cite{pushshiftchange}. 
The lack of clarity or guidelines by Reddit regarding API access applications for research purposes further exacerbates this issue.

\subsection{Future Work}
In this work, we further explored the complexities of the VTuber ecosystem beyond existing research. 
As demonstrated in previous sections, the use of avatars by online content creators has deep implications on how they create, maintain and switch online identities (e.g., after retiring a previous VTuber identity),  corporate brand management and viewer dynamics with other viewers, content creators and corporations.
We belief that many of these implications are also applicable more generally to research areas like embodiment, social media and online identities, fandoms, and parasocial relationships. We highlight a few of these here: 

\textit{Parasocial relationships.} Although we observed evidence of a portion of viewers being aware of their parasociality, it is unclear how they moderate (if at all) their parasociality. Do different online spaces where viewers gather serve to moderate or amplify parasociality? Do viewers grieve different types of content creators (e.g., VTubers vs. non-VTubers, live-streamers vs. video-on-demand creators, popular vs. less popular creators) differently? 

\textit{Embodiment.} While prior VTuber research found no significant preference for either 2D or 3D avatars besides their differing degrees of expressiveness \cite{Lu2021}, it is unclear how the affordance of different types of embodiment (e.g., different degrees of freedom and customizability) influence the spectrum of content that can be created. E.g., having a 3D model could allow VTubers and viewers to share a single 3D VR space, having customizable avatar lighting allows viewers to be more immersed when an avatar's lighting matches the game during Let's Plays.
It is also unclear if different avatar types could lead to different parasocial dynamics or grieving behaviors.

\textit{Online Identities.} We did not explore the simultaneous use of multiple identities, e.g., if a nakanohito creates content via two VTuber identities concurrently or only chooses to use an avatar for certain types of content. On one hand are questions regarding how content creators manage this, while on the other hand are possible explorations into how viewers navigate their perceptions of the different concurrent identities of a single creator.

\section{Conclusion}

{Given the significant increase in VTuber viewership and the retirement of several popular VTubers}, we investigate English-speaking viewers' reactions when parasocially grieving retired VTubers. 
We identified relevant subreddits and the two most discussed VTuber retirements on Reddit through an exploratory phase: Kiryu Coco and Uruha Rushia of hololive production.
Based on the proposed typology of VTuber retirements, Coco's graduation and Rushia's termination provided grounds for comparing different types of announced retirements.
Posts and comments from relevant subreddits spanning nearly three years were collected and filtered based on the inclusion criteria.
The filtered dataset of 13655 posts and comments was then analyzed using mixed methods.
Our analysis found that most negative emotions dissipated over time, with Coco's graduation eliciting reactions that were less negative than Rushia's termination.
The use of both known and novel coping methods was observed, in addition to grief policing behaviours that stemmed from a strong shared identity.
Moreover, discussions of VTuber reincarnations revealed the role of interactions between content creators in the construction of a VTuber's identity.
Differences in viewer behaviours across the analyzed subreddits due to the subreddits' contrasting characteristics were also discovered.
Finally, we discuss how the processes of parasocial grieving and tribute creation can be better supported with several design implications, alongside {future research directions}.


\bibliographystyle{ACM-Reference-Format}
\bibliography{main}


\appendix
\newpage
\section{Introducing Kiryu Coco and Uruha Rushia}
\label{app:vtuberintro}
\subsection{Kiryu Coco}
Kiryu Coco debuted on December 28, 2019 \cite{cocohololivepage}, just a few months before pandemic lockdowns were implemented in major parts of the world. 
Lore-wise, Coco was a dragon from another world who went to Japan as a language exchange student. 
Inspired by her love of the Yakuza game series, viewers also often refer to her as the \textit{kaichou}, meaning chairman in English, of a fictional Japanese gang.
Her fans were referred to as \textit{kiryukai}.
The first generation of hololive English, Myth, would not launch till mid-September 2020. As such, Coco, who was proficient in both Japanese and English, attracted many {English-speaking} viewers. Moreover, Coco actively engaged the r/Hololive subreddit community through her live stream series, named \textit{reddit [expletive]post review!} \cite{cocomemereviewpost,cocomemereviewplaylist}, in which Coco invited other hololive talents to join her in reviewing memes and other submissions on the subreddit. 
Coco was also known for \textit{AsaCoco Live News}, a satirical news program in which she presented hololive-related news \cite{cocoasacocoplaylist}, often starting the stream with her signature English greeting: ``Good morning, [expletive]!''.

Such initiatives led to Coco having a sizeable audience. As of the time of writing, more than two years after her graduation, Coco's YouTube channel boasts 1.39 million subscribers \cite{cocoyoutubechannel} and almost USD\$3 million dollars in superchats \cite{playboardsuperchat}. 
Coco's retirement was officially announced by hololive on June 9, 2021, and she held her graduation live stream on July 1, 2021. The live stream included chats with other hololive talents and a virtual concert with her fellow forth-gen talents.
Her videos and live stream archives are still available on her YouTube channel \cite{cocoyoutubechannel}.

\subsection{Uruha Rushia}
Uruha Rushia debuted on July 18, 2019 \cite{uruharushiafandom}, whose character was one of a necromancer, with her fans known as \textit{fandeads}. 
She was known for occasionally raging while playing games, being possessive towards her viewers, feeling jealous when her viewers bring up other females during her streams (e.g., \cite{rushiajealous,rushiapossessive,rushiaangry1,rushiaangry2}), and playing a character who was self-conscious about some of her avatar's physical features \cite{rushiajokecontext}.

Rushia's audience size was similar to Coco's, with 1.36 million YouTube subscribers \cite{rushiayoutubechannel} and more than USD\$3 million dollars in superchats \cite{playboardsuperchat}. 
On February 24, 2022, hololive announced their immediate termination of Rushia due to reasons related to contract breaching \cite{rushiaRedditAnnouncement}. 
Unlike Coco, all of Rushia's YouTube videos and live stream archives were either removed from public access or deleted around a month after her termination \cite{rushiayoutubechannel}, and there was no final farewell live stream.

\onecolumn

\section{Data Collection Details}

\begin{table}[h]
\caption{Details of the data collected using each VTuber's name and nicknames as search terms. For each subreddit, multiple search terms could have retrieved identical posts, as such, the total number of unique posts is less than the total number of collected posts.}
\label{tab:datacomp}
\begin{tabular}{cccccc}
   \makecell{VTuber \\(Total posts/comments)} & Subreddit & Search Term & \#Searched Posts & \#Posts & \#Top-level Comments \\
  \hline
  \multirow{25}{*}{\makecell{Kiryu Coco\\(1407/21676)}} & \multirow{12}{*}{r/Hololive} & Kiryu & 129 & \multirow{12}{*}{675} & \multirow{12}{*}{19446}\\
  && Coco & 97 &\\
  && Kiryu Coco & 126 &\\
  && Kaichou & 94 &\\
  && Chairman & 25 &\\
  && Coco-chin & 3 &\\
  && Coco-chi & 1 &\\
  && Yakuza Dragon & 116 &\\
  && Kiryu-chan & 74 &\\
  && Yuujin C & 5 &\\
  && Friend C & 56 &\\
  && C-chan & 109 &\\
  \cline{2-6}
  & \multirow{12}{*}{r/VirtualYoutubers} & Kiryu & 120 & \multirow{12}{*}{424} & \multirow{12}{*}{1776}\\
  && Coco & 119 &\\
  && Kiryu Coco & 107 &\\
  && Kaichou & 66 &\\
  && Chairman & 9 &\\
  && Coco-chin & 0 &\\
  && Coco-chi & 0 &\\
  && Yakuza Dragon & 43 &\\
  && Kiryu-chan & 7 &\\
  && Yuujin C & 0 &\\
  && Friend C & 86 &\\
  && C-chan & 34 &\\
  \cline{2-6}
  & r/KiryuCoco & \textit{None used} & 308 & 308 & 454\\
  \hline
  \multirow{13}{*}{\makecell{Uruha Rushia\\(636/6499)}} & \multirow{6}{*}{r/Hololive} & Uruha & 29 & \multirow{6}{*}{101} & \multirow{6}{*}{3791} \\
  && Rushia & 12 &\\
  && Uruha Rushia & 28 &\\
  && Rushifer & 2 &\\
  && Boing Boing Rushia & 19 &\\
  && Cutting Board & 41 &\\
  \cline{2-6}
  & \multirow{6}{*}{r/VirtualYoutubers} & Uruha & 73 & \multirow{6}{*}{242} & \multirow{6}{*}{1408}\\
  && Rushia & 62 &\\
  && Uruha Rushia & 61 &\\
  && Rushifer & 1 &\\
  && Boing Boing Rushia & 137 &\\
  && Cutting Board & 3 &\\
  \cline{2-6}
  & r/Rushia & \textit{None used} & 293 & 293 & 1300\\
  \hline
  Total & & & & 2043 & 28175 \\
  \hline
\end{tabular}
\end{table}

\newpage
\section{Inductive Codebook}
\label{app:codebook}

\begin{table*}[htbp!]
\caption{Codebook with the inductive codes used during qualitative analysis, in addition to the deductive codes presented in Table \ref{tab:deductivecodes}.}
\label{tab:codebook}
\begin{tabular}{p{2cm}p{5cm}p{8cm}}
   Code & Definition & Examples\\
\hline
   Emotion:\newline Respect & Showing respect to the retiring/retired VTuber & o7\newline The news hurts but I respect Kaichou's decision\\
\hline
   Emotion:\newline Loyalty & Being dedicated to remembering and supporting a VTuber or nakanohito post-retirement & Kiryukai forever\newline I will kneel forever\newline I'll always be supporting her\\
\hline
   Emotion:\newline Concerned & Being worried about a VTuber's nakanohito & We are worried about you \newline I cannot help but be concerned for her mental health\\
\hline
   Emotion:\newline Regret & Feeling apologetic over not having done something prior to a VTuber’s retirement or its announcement & I am sorry for not watching your streams\newline I missed the stream T-T\newline I regret not continuing her channel membership\\
\hline
   Coping:\newline Merchandise & Discussions around a retiring/retired VTuber's merchandise or sharing one's purchased merchandise & Hopefully I can buy her birthday merchandise\newline \textit{Other examples include sharing photos of merchandise like phone cases, pins, plushies} \\
\hline
   Coping:\newline Hypothetical Situations & Imagining hypothetical situations for a retiring/retired VTuber & Not surprised if she becomes a manager for other VTubers \\
\hline
   Coping:\newline References & Sharing content that contains mentions of/references to a retiring/retired VTuber & \textit{Examples include sharing tweets by other influencers that refer to a retired VTuber or live streams by other VTubers in which they talked about a retired VTuber}\\
\hline
   Coping: Refusal\newline to Leave & Staying at an online space and refusing to leave it  & A screenshot of her videos before they are deleted\newline 10 hours before her YouTube channel is terminated\newline I'M NEVER CLOSING THIS TAB\\
\hline
   Coping:\newline Substitutes & Finding substitute VTubers or content to watch & Is anyone else taking over meme reviews?\newline Are there any VTubers like Coco?\\
\hline
   Coping: Taking\newline a Break & Taking a break from watching VTubers or social media & I'm gonna refresh myself in the outside world for some time\newline I will take a break\newline I don't want to browse anymore\newline This will be the final day of me watching VTubers\\
\hline
   Reincarnation & Mentions of, or discussions about, other VTuber identity/identities embodied by a retired VTuber’s nakanohito post-retirement & I hope she goes indie\newline Your fans will find you again in their own way\newline See you next life\newline Will she stream on the alt account?\\
\hline
   Policing & Norm enforcement practices around grief \cite{Gach2017} & Make sure to not discuss this in other VTubers' streams\newline I want folks to be understanding\newline Do be reasonable\\
\hline
   Archival & Discussions and questions about the archival status of a retiring/retired VTuber’s content, including whether a VTuber’s videos will remain public on their social media or about creating backup copies of their videos  & Are her old streams going to still be up in the future?\newline Don't know if downloading her video is legal but I want a copy\newline Are we allowed to archive her videos?\newline Please do not make her channel private\newline Membership stream archive should not be shared publicly\\
\hline
   Collaboration & Discussions about viewers working together or sharing outcomes of collaborative processes & Sharing a short movie by Tatsunokos, with love\newline We collected over 1000 messages, 300 fanarts from many fans\newline Thanks to every single person who helped make this video\\
\hline
\end{tabular}
\end{table*}


\end{CJK}

\end{document}